\documentclass[aps,prb,onecolumn,floats]{revtex4-2}
\usepackage{epsfig}
\usepackage{amsmath, amssymb}
\usepackage{graphicx}
\usepackage{soul}
\usepackage{braket}
\usepackage[colorlinks=true,linkcolor=blue]{hyperref}
\usepackage{subfigure}
\usepackage{color}

\begin{document}
\title{Topology and Localizations in a 2D Su-Schrieffer-Heeger Model with Domain Walls, Quasi-periodic Disorder and Periodic Hopping Modulations}
\author{Surajit Mandal$^{1,2}$}
\email{surajitmandalju@gmail.com}
\author{Satyaki Kar$^{2}$}
\email{satyaki.phys@gmail.com}
\affiliation{$^{1}$Department of Physics, Jadavpur University, Kolkata, West Bengal -700032, India\\$^{2}$Department of Physics, AKPC Mahavidyalaya, Bengai, West Bengal -712611, India}
\begin{abstract}
 { We study a two dimensional (2D) Su-Schrieffer-Heeger (SSH) model on a square lattice in presence of domain walls (DW) / vortices or quasi-periodic disorders to investigate the nature of topology and localizations in its quantum states.
While in a pure 2D SSH model, zero energy states (ZES) lie within the dispersion continuum and the bound states in continuum (BIC) are localized at the corners, a continuous distributions of DWs can produce localized ZES along the DW lines or at the DW center depending on the orientations of the DWs. Moreover with such DWs, one can witness nonzero energy in-gap states showing localizations at the edges, along the DWs or at the DW center. For probing disorder effect, we introduce on-site quasiperiodic potentials (QP) in such systems that show the usual tendency of the states to localize. But exotic reentrant localization behavior is also captured for judicious choice of the QP term. We also examine the scenario for different hopping periodicities in the SSH Hamiltonian. Interestingly for anisotropic hopping modulations, the bulk ZES gets exhausted leaving only topological boundary modes at zero energies. The fate of these states in presence of the DWs are also discussed. Our present study with its plethora of exotic outcomes can thus inspire varied applications in the field of topological quantum computations. }
\end{abstract}
\maketitle
\section{Introduction}\label{sec1}
 Over the last couple of decades, the study of topological insulators (TI) has seen genuine enthusiasm from the condensed matter community 
 for the robustness they show in their conducting edge/surface states\cite{kane,zhang1}.
Tuning of relevant system parameters in these materials {results in transition from} the trivial non-topological phases (NTP) to topological phases (TP) {where a bulk-boundary correspondence connects the topological invariant of TP with its robust boundary modes}. Normally, this bulk-boundary correspondence refers to the first-order $d$-dimensional TIs, where the $d$-dimensional ($dD$) system yields $dD$ gapped bulk states and $(d-1)D$ gapless boundary states. In recent seminal works\cite{hughes,edge}, such study has been generalized further to include {second\cite{second} or} higher-order topological insulators (HOTIs), which probe lower-dimensional gapless boundary states, $i.e.$ for a $dD$ TI ($n$th-order TIs) one can now have $(d-n+1)D$ gapped boundary states and $(d-n)D$ gapless boundary states for $n\geq 2$\cite{hughes,edge}. 

In this respect, nearest neighbor 1D hopping model with Su-Schrieffer-Heeger type hopping alternations\cite{su} has acquired quite a fame as toy model for TI as well as the system of carbon chains in Polyacetylene\cite{heeger}. Particularly, a 2D version of it has already shown the presence of higher order topological modes\cite{edge}.
 Jackiw and Rebbi\cite{rebbi} and Su, Schrieffer, and Heeger\cite{su} showed that states with fractionalized charge of $e/2$ are localized at a domain wall separating two degenerate ground states in Polyacetylene\cite{heeger}. This results from lattice dimerization and the intermediate domain wall where a zero-mode with fractional charge interpolates between two degenerate vacua. Such charge fractionalization implies differences of the bulk charge polarization of these two degenerate ground states.

Of late, the study of a two dimensional (2D) version of this SSH type bond alternations has gained momentum for its exotic topology, especially for being a candidate HOTI\cite{lee,liu,kar,karc}.
Recently Hou, Chamon, and Mudry (HCM) considered a graphene-like system to realize the idea of fractional charges in two dimensions ($2D$)\cite{hou}. Their study reveals that a vortex in the order parameter for the Kekul\'{e}-type dimerization guarantees the presence of zero-energy states (ZES) with fractional charge. Later, the study performed in Ref.\cite{lee} finds that the ground states of Kekul\'{e}-textured graphene, similar to $1D$ SSH chain, are characterized by a $\mathbb{Z}_{2}$ topological invariant $\nu_{2D}$.

Motivated by these findings of charge fractionalizations observed at the location of topological defects and realizing that the reported studies of topological 2D SSH models lack similar important extensions, we introduce defects via domain walls or vortices in such a model in order to investigate the nature of ZES at the defect locations as well as the charge fractionalizations there, if any. In particular, we {consider} a series of domain walls distributed in two different manner - the first one featuring two orthogonal DW lines intersecting at the center of the lattice while the second one gives closed circular DW loops arising from radially symmetric hopping modulations.
{ As periodic hopping modulations enriches considerably the spectra and topology in a SSH chain\cite{kar}, we also incorporate the same in our 2D model with defects.
Moreover, as} disorder comes naturally in real physical systems, for completeness we investigate the outcome of disorders as well in the 2D SSH model. Disorder plays an important role in shaping the localization scenario of the wavefunctions.
In an Aubry-Andre (AA) model\cite{aubry}, it is treated not randomly like in an Anderson model\cite{ander} but with a quasi-periodic potential with a spatial period that is incommensurate with the lattice period. Such model is crucial for studying localization phenomena as it motivates disorder driven metal-insulator transitions in lower dimensional systems. {Recently a 1D SSH chain, subject to such disorders, has been shown to produce re-entrant localization behavior\cite{s-basu}. Intrigued by such finding,} we introduce such quasiperiodic disorders in our 2D extended SSH model to find out how localization build up there both in the topological and trivial states. Interestingly, our results also indicate re-entrant localization behavior on application of such disorder {of suitable choice and parameter regime}.

The paper is organized as follows. {We start with the formulation of the $2D$ SSH model in the Section II. It also discusses many hitherto overlooked minute details} of edge and corner states for various dimerization strengths. Section III discusses the implication of introducing domain walls of different kinds in such systems. There the numerical results for ZES and in-gap states are analyzed in detail. The effect of quasiperiodic disorders is discussed in the following section IV. In section V, we consider other commensurate variation of hopping periodicity given by $\theta_x=\theta_y=\pi/2$ and $\theta_x=\pi,~\theta_y=\pi/2$ (to be defined later) and provides a comparative analysis of them with the usual 2D SSH hopping variations. Finally, in Section VI, we summarize the findings to conclude our work and mention about the practicability of the problem and its future possibilities.

{\section{2D SSH Model: Spectra \& Topology}}\label{sec3}
With symmetric hopping\cite{liu}, a 2D SSH Hamiltonian in a $L\times L$ square lattice under open boundary condition (OBC) can be written as
\begin{equation}\label{1}
H=\sum_{i,j}^{L-1,L}(t+\delta_{i})c_{i,j}^{\dagger}c_{i+1,j}+\sum_{i,j}^{L,L-1}(t+\delta_{j}^{\prime})c_{i,j}^{\dagger}c_{i,j+1}+hc,
\end{equation}
{where $t$ is the nearest neighbor hopping parameter and $\delta_{i}=\Delta\cos[(i-1)\theta_{x}]$ and $\delta_{j}^{\prime}=\Delta\cos[(j-1)\theta_{y}]$ denote hopping modulations in the nearest neighbor bonds along $x$ and $y$ directions respectively from the site with index $(i,j)$ in a square lattice. Thus $\Delta$ represents the amplitude of the periodic hopping modulations. Besides,} $c_{i,j}^{\dagger}(c_{i,j})$ denotes the creation (annihilation) operator of site ($i,~j$). Notice that for a periodic system, all the upper-limits in the summations of indices in Eq.\ref{1} will be $L$. In this section, we perform a detailed analysis of the spectral and topological features of this model{, both under periodic and open boundaries for $(\theta_x,\theta_y)=(\pi,\pi)$}. Notice that in a $L\times L$ square lattice {with periodic boundaries}, such Hamiltonian produces $L^2/4$ number of 4-site unit cells with total number of sites being $N=L^{2}$. {In other words, there are 4 sublattices.}\\

\begin{figure}
   \vskip 0 in
   \begin{picture}(100,100)
     \put(-180,0){
     \includegraphics[width=.4\linewidth,height= 1.5 in]{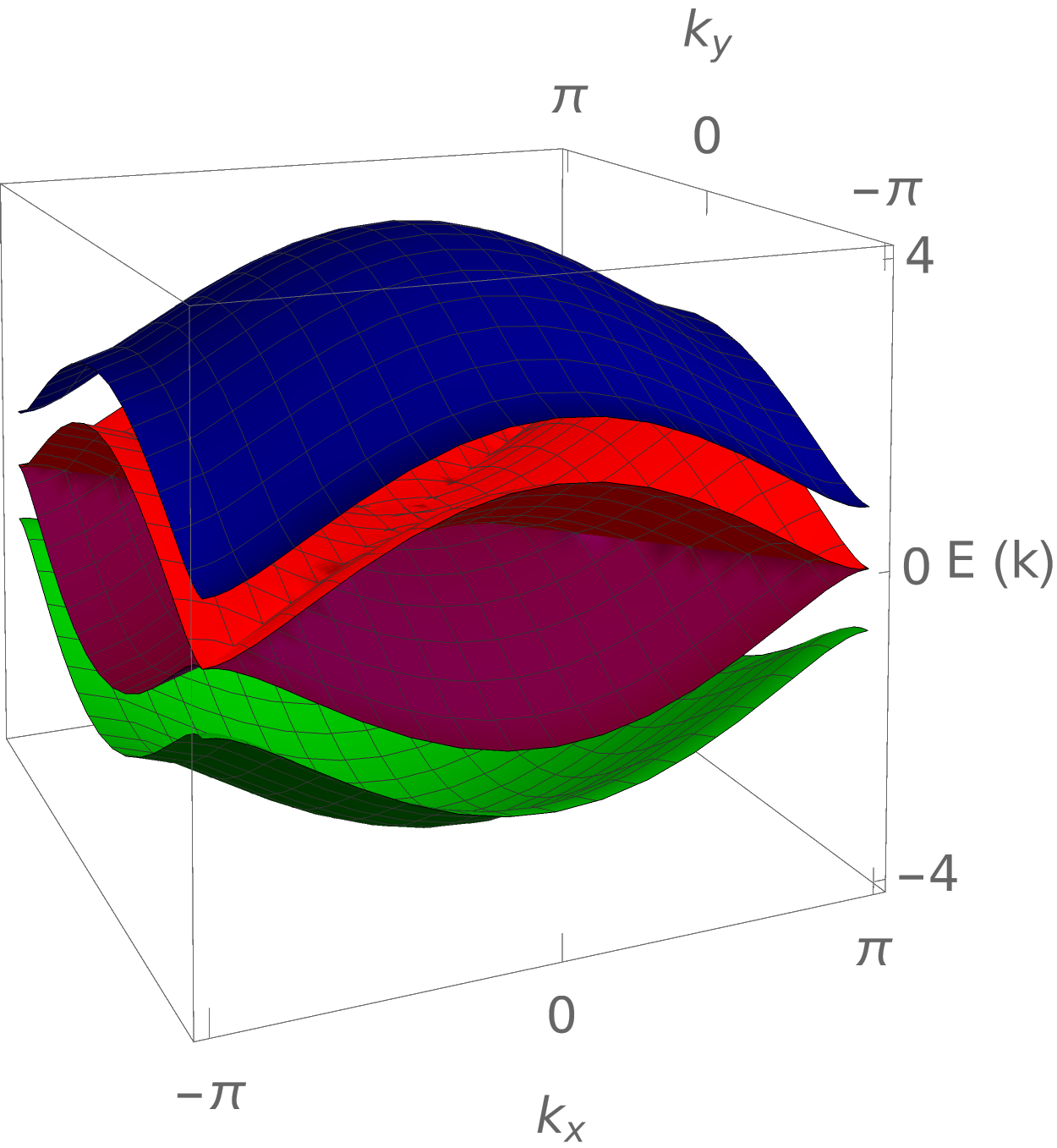}\hskip .4 in
  \includegraphics[width=.4\linewidth,height=1.5 in]{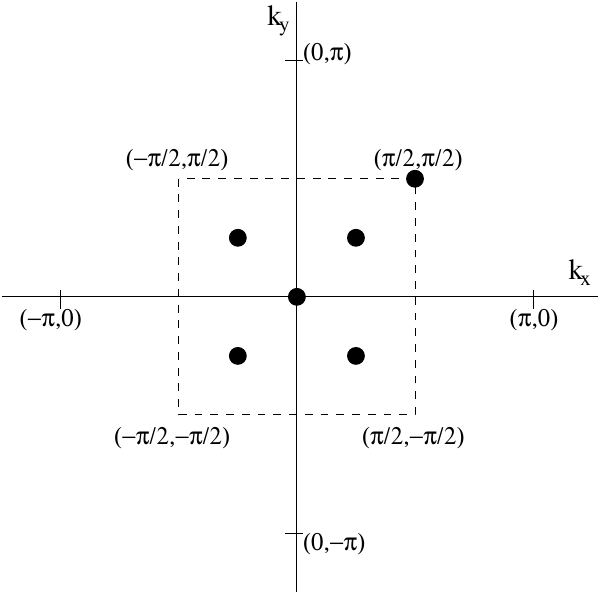}}
     \put(-150,100){(a)}
     \put(76,100){(b)}
   \end{picture}\\
   \vskip .2 in
   \begin{picture}(100,100)
     \put(-150,-10){
       \hskip -.1 in \includegraphics[width=.42\linewidth,height=1.5 in]{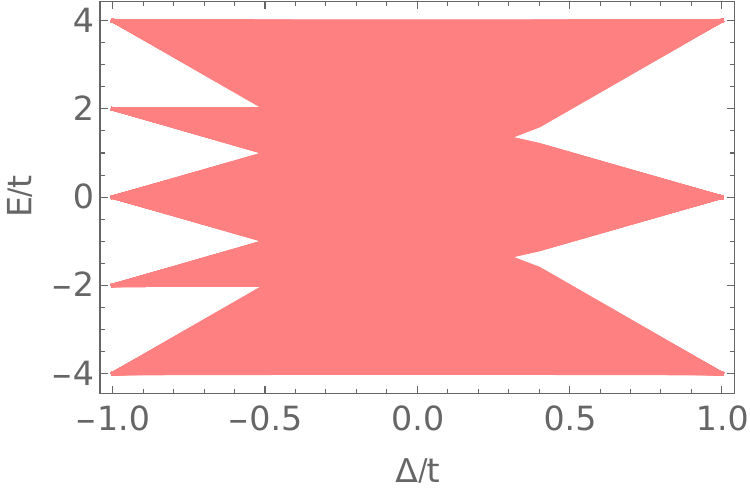}
       \hskip 1 in
       \includegraphics[width=.3\linewidth,height=.65 in]{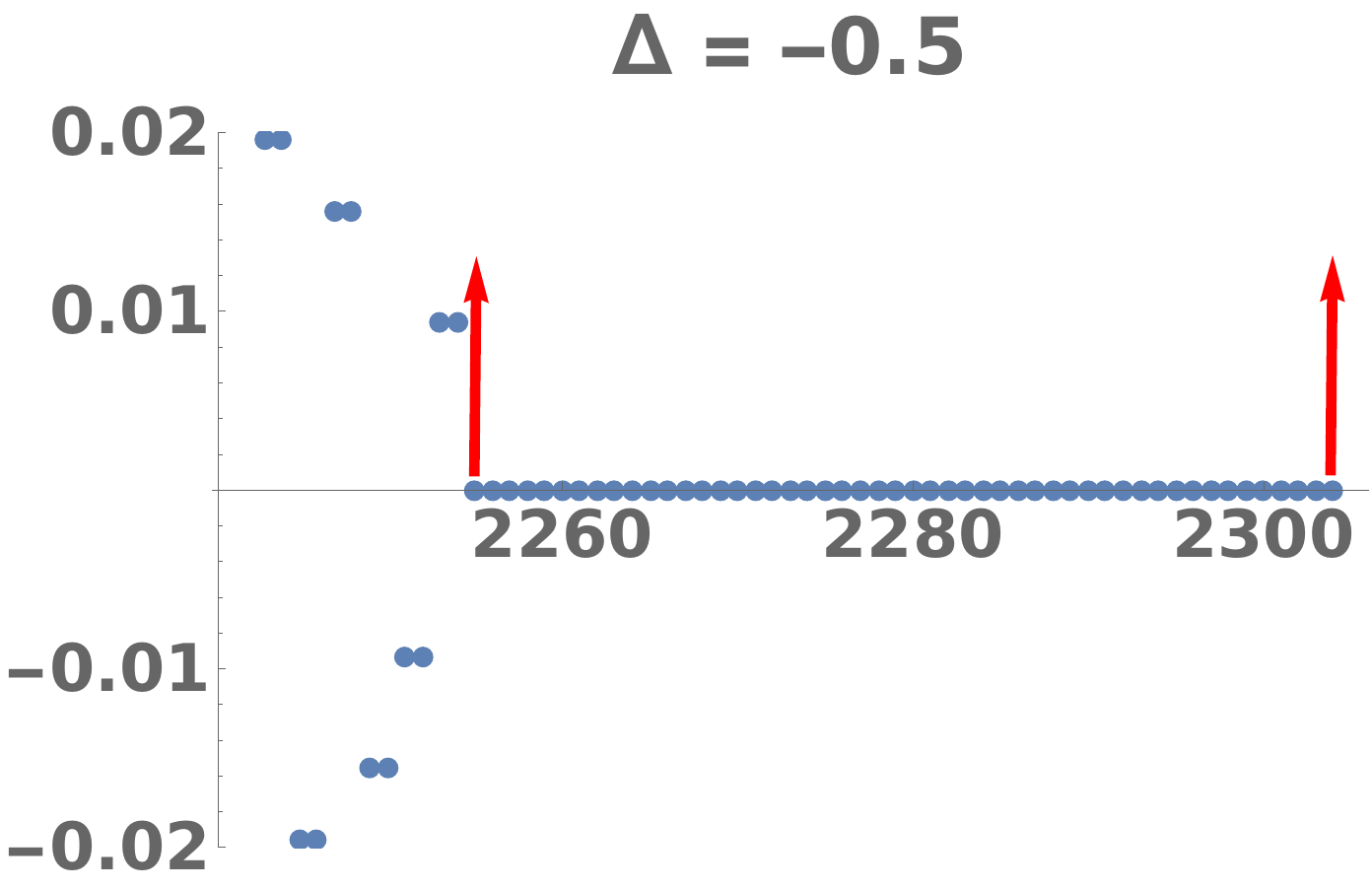}}
     \put(-140,110){(c)}
     \put(145,105){E}
   \put(185,94){\tiny{2257}}
   \put(280,94){\tiny{2304}}
   \put(255,110){(d)}
   \put(215,60){n}
   \put(145,35){E}
   \put(180,25){\tiny{2255}}
   \put(280,25){\tiny{2304}}
    \put(255,40){(e)}
    \put(215,-10){n}
  \end{picture}
   \begin{picture}(100,100)
     \put(30,60){
             \includegraphics[width=.3\linewidth,height=.65 in]{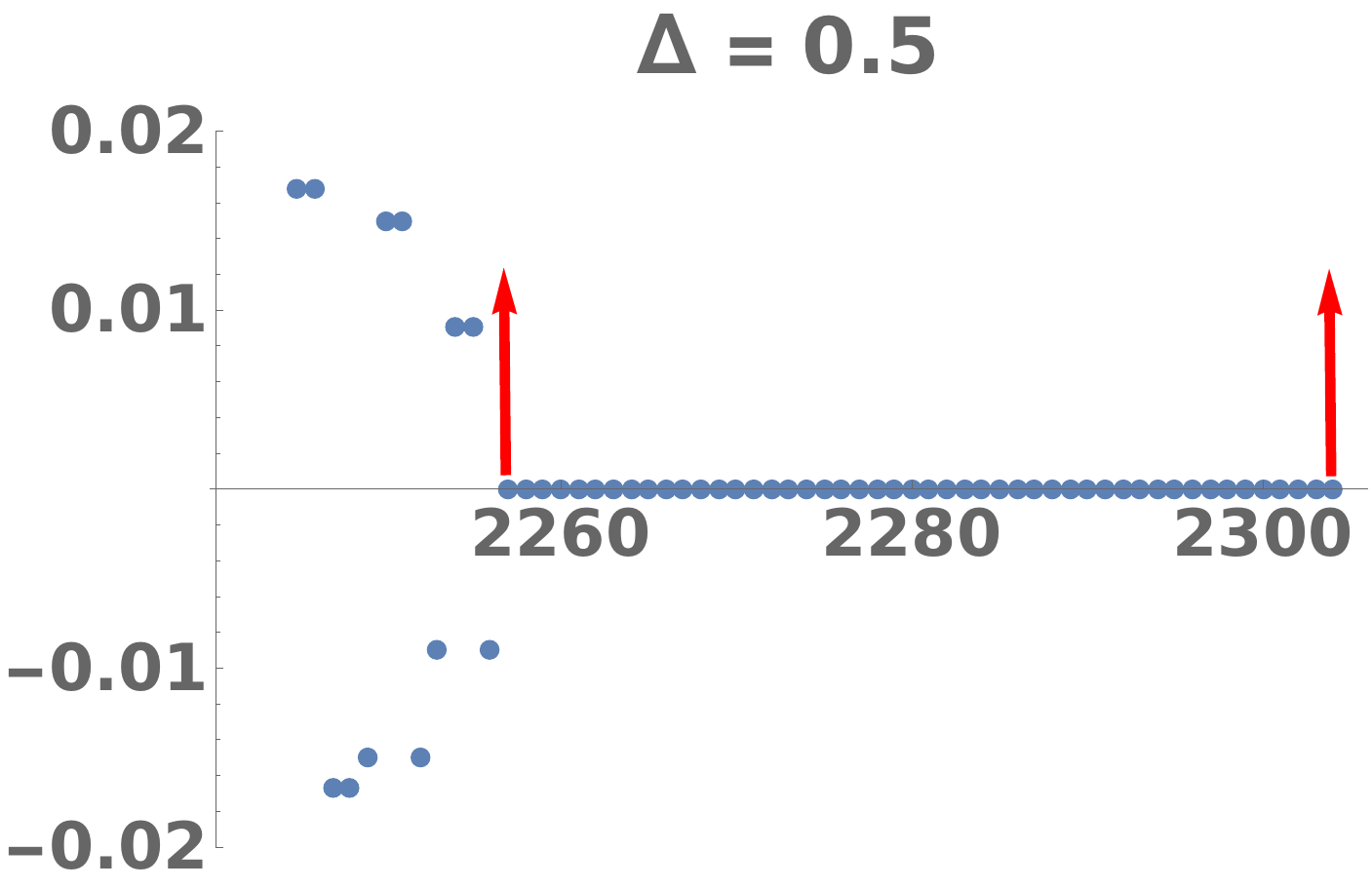}}
 \end{picture}

  \vskip 0.1 in
\caption{(a) Typical band structure of a $2D$ SSH model under PBC as a function of $k_{x}$ and $k_{y}$ away from the TQPT. (b) $L-2$ number of points with $|k_x|=|k_y|$ within the BZ in a $L\times L$ square lattice (with $L=8$) where doubly degenerate zero energy modes appear. (c) Dispersions from a $48\times48$ $2D$ SSH model under OBC of which (d) and (e) demonstrate the $L$ and $L+2$ number of ZES in typical trivial and topological phase respectively ($n$ being the eigenstate indices {and $t=1$}).} 
\label{fig1}
\end{figure}

{\subsection{System with Periodic Boundary}}
For the case of $\theta_{x} = \theta_{y} = \theta = \pi$, {one can Fourier transform the above Hamiltonian to get a 
 $4\times 4$ Bloch Hamiltonian $H_k$ (see Appendix \ref{ap1})} in the momentum
space. This results in 4-band dispersions as shown in Fig.\ref{fig1}(a) for typical $\Delta/t$ values away from the topological quantum phase transition (TQPT) point.
The spectrum is symmetric around zero energy due to chiral symmetry.
 Moreover, it also respects $C_{4v}$ symmetry\cite{chiral,liu}{, as it remains invariant under a $\pi/2$ rotation about the {perpendicular} $\hat z$ axis. This makes spectrum $E(k_x,k_y)$ to be functions of only the magnitudes of the momentum:
  $|k_x|$ and $|k_y|$. But from the Block-antidiagonal form of $H_k$ in which it can be transformed into (see Appendix \ref{ap1}), one can readily see that $|k_x|=|k_y|$ leads to zero energy eigenvalues. Thus} the simultaneous presence of chiral and $C_{4v}$ symmetries lead the lattice to always have zero energy gapless bulk energy bands\cite{Benalcazar}. In general, the dispersion for a periodic system looks like:
\begin{widetext}
\begin{equation}\label{4}
E(k_{x},k_{y})=\pm \Bigg[4(t^2+\Delta^2)+2(t^2-\Delta^2)(cosk_{x}+cosk_{y})\\ \pm 8\sqrt{\Big(t^2 cos^2(k_{x}/2)+\Delta^2 sin^2(k_{x}/2)\Big)\Big(t^2 cos^2(k_{y}/2)+\Delta^2 sin^2(k_{y}/2)\Big)}\Bigg]^{\frac{1}{2}}  
\end{equation}
\end{widetext}

{It indicates that the model under periodic boundary condition (PBC) possess} two pairs of particle-hole symmetric dispersion bands with no energy gap between the low energy pair along the nodal direction $|k_{x}|=|k_{y}|$ (notice Fig.\ref{fig1}(a)). Particularly, the band gap between all four bands vanishes at $(k_{x},k_{y})$=($\pm\pi,\pm\pi$) for the isotropic point $\Delta/t=0$. 
At this isotropic point, band inversion {(see Appendix \ref{ap1})} occurs at some high symmetry points (HSP) of the BZ, namely at X(Y) points ($\pi$,0)~((0,$\pi$)) while band crossing (without parity inversions) occur at the M ($\pi,\pi$) point {(see Appendix \ref{ap1} for details)}.
Thus at X(Y) point, a TQPT occurs and this also brings in nonzero fractional polarizations in the topological regime of $\Delta/t<0$\cite{liu}.\\

{\subsection{Zak Phases and ZES}}

{Although the bulk spectrum is gapless along the nodal lines, one can define a vector Zak phase $Z(Z_x,Z_y)$ in this system as the Hamiltonian {enjoys dimensionality reduction for $\theta_x=\theta_y,~i.e.,$} it is separable and can be written as $H(k_x,k_y)=H_x(k_x)\otimes \hat{I_2}+\hat{I_2}\otimes H_y(k_y)$, $H_x$ and $H_y$ being 1D SSH Hamiltonians along $x$ and $y$ directions respectively.} The $l^{th}$ component of such a Zak phase can be calculated as 
\begin{equation}\label{zak}
  Z_l=i\sum_{j=1}^{N_{occ}}\oint_{BZ}\langle u_j(k)|\frac{\partial}{\partial k_l}|u_j(k)\rangle dk_l
\end{equation}
$|u_j(k)\rangle$ and $N_{occ}$ being the $j$-th eigenstate and the no. of occupied bands\cite{lee,liu} respectively. {Notice that for any fixed $k_x(k_y)$, the integration along $k_y(k_x)$ encounter only gapped phases barring at the point $|k_x|=|k_y|$ and is independent of $k_x(k_y)$. We find} $Z=(\pi,\pi)$ (see detailed calculations in the Appendix \ref{ap1}) in the topological phases where band inversions occur at the $X(Y)$ point. Like in 1D, the nonzero Zak phase correspond to in-gap edge states in systems with OBC though they no more remain zero energy mid-gap states. Fractional polarization is also realized as $Z$ is related to the polarization as $P_l=\frac{Z_l}{2\pi}$\cite{obana} yielding $P=(\frac{1}{2},\frac{1}{2})$ in the topological phase. Physically, this corresponds to a mismatch between ionic and electronic charge distributions, leading to fractional boundary charges in a finite system. Such fractionalization is protected by crystalline (reflection and $C_{4v}$) symmetries, not by a bulk spectral gap.
 Unlike the insulating (bulk-insulating) nature of the topologically trivial (non-trivial) regime of $\Delta/t<0$ ($\Delta/t>0$) in the $1D$ system\cite{kar,mandal,nhmandal}, this 2D system features a metallic ({or more precisely, a nodal-line semimetallic}) phase in both situations due to the gaplessness along $|k_x|=|k_y|$. In fact, one can find the number of ZES by counting all possible modes along the $-M-\Gamma-M$ and $-M'-\Gamma-M'$ directions (with $M'=(-\pi,\pi)$) that falls within the 1st Brilluin zone (BZ) of the 4-sublattice structure (bounded by corners $[\pm\frac{\pi}{2},\pm\frac{\pi}{2}]$), each with a degeneracy of 2. In a 2D SSH periodic $L\times L$ square lattice, that number comes out to be $2L-4$ (see Fig.\ref{fig1}(b)), for any $\Delta$ away from the TQPT. 
\\

{\subsection{System with Open Boundary}}

The open boundary condition brings in edge states (as well as corner {localized} states) in the topological phases. Interestingly, here one encounters a special band structure (see Fig.\ref{fig1}(c)) where the topological zero energy modes appear within the bulk spectral continuum (and not within a spectral gap), and {thus they remain} degenerate with a number of bulk modes\cite{wei}.
In fact in a $L\times L$ lattice with open boundaries, there are $L$ number of zero energy bulk modes in the trivial regime ($eg$, see Fig.\ref{fig1}(d)) while $L+2$ number of zero modes deep within the topological regime(see Fig.\ref{fig1}(e)).
So two nonzero energy bulk modes of the trivial regime turn into corner modes within the topological regime. Two zero energy bulk modes also turn into BICs at the same time.
\\

{\subsubsection{Corner-localized BICs}}

\begin{figure*}
     \includegraphics[width=.249\linewidth]{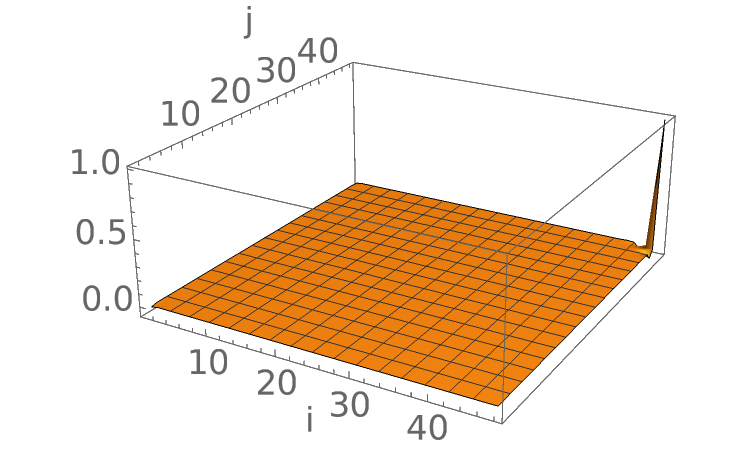}
  \includegraphics[width=.24\linewidth]{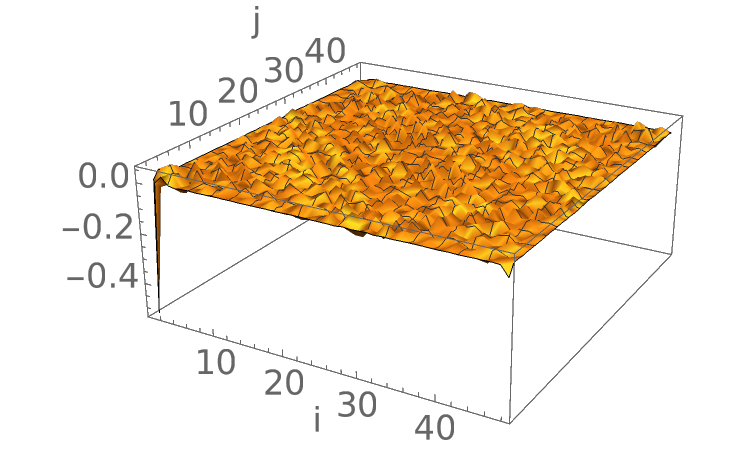}
  \includegraphics[width=.24\linewidth]{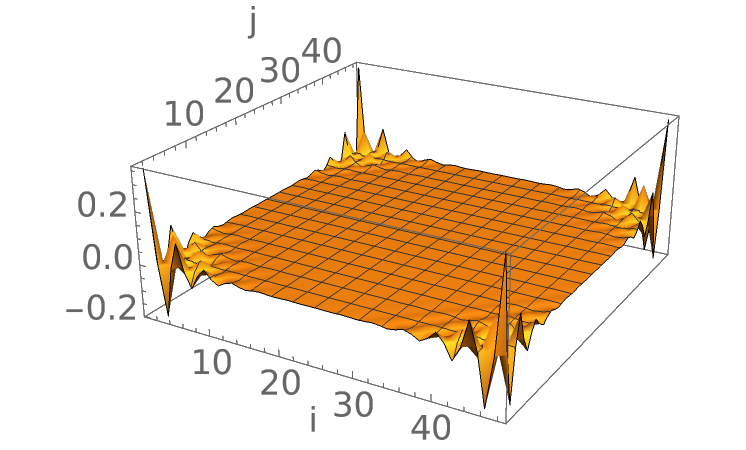}
  \includegraphics[width=.24\linewidth]{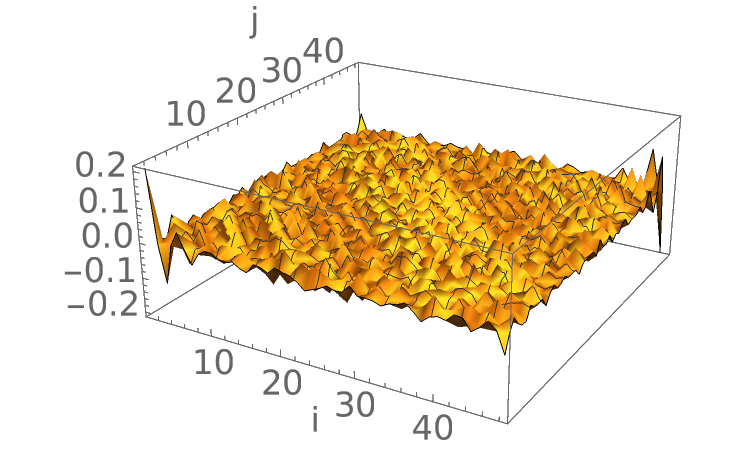}
     \put(-430,68){(a)}
     \put(-300,68){(b)}
     \put(-170,68){(c)}
     \put(-50,68){(d)}\\
   \includegraphics[width=.24\linewidth]{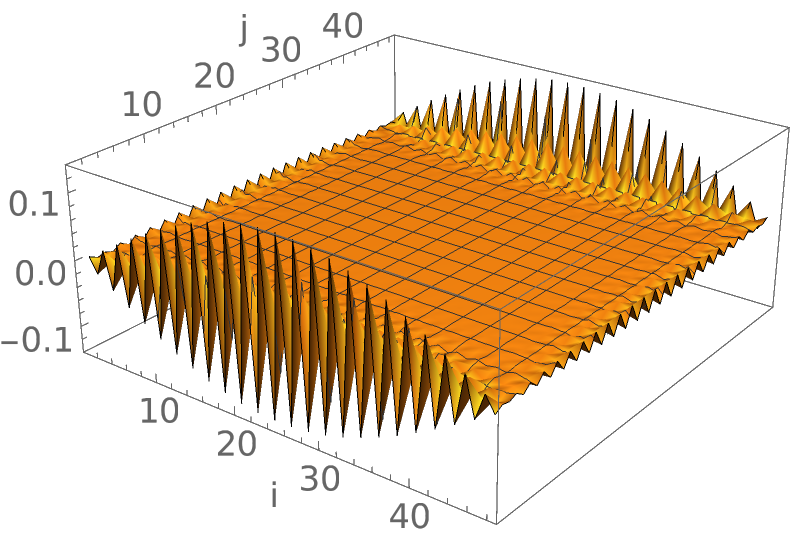}
  \includegraphics[width=.24\linewidth]{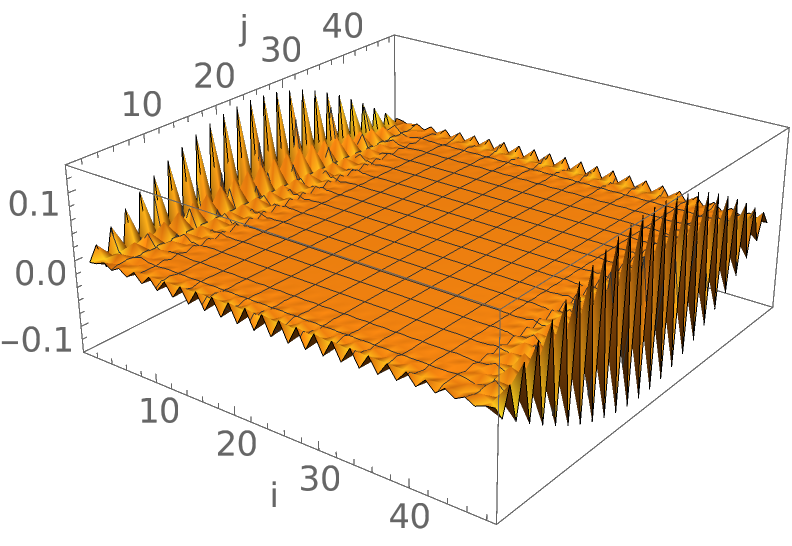}
  \includegraphics[width=.24\linewidth]{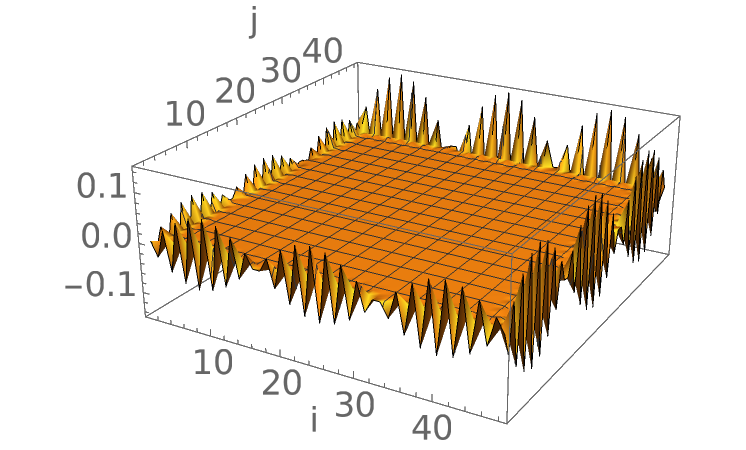}
       \put(-300,76){(e)}
     \put(-176,76){(f)}
   \put(-50,77){(g)}
\caption{{\bf Zero energy} corner localized BIC's [(a),(c)] and  BIC's mixed with bulk states [(b),(d)] in a $2D$ SSH model in a $48\times 48$ square lattice for $\Delta/t$ = - 0.9 (a-b) and -0.3 (c-d) respectively. (e-g) show {\bf nonzero energy} in-gap states for $\Delta/t=-0.5$ with localization along (e) $x$-edges, (f) $y$-edges and (g) both $x$ and $y$ edges respectively.} 
\label{fig2}
\end{figure*}

The zero energy bound states within the bulk continuum that coexist with degenerate bulk modes at zero energy, are commonly called bound states in the continuum (BICs)\cite{Benalcazar}.
In the configuration I of Ref.\cite{ma}, which is also the outcome of our Hamiltonian for $\Delta/t<0$, there are four nearest-neighbor bonds with stronger hopping constituting bonding squares that cover all the inner sites while the sites at the four corners remain isolated with weaker hopping alone. This gives rise to corner-localized BICs whose penetrations from all the corners into the bulk are suppressed exponentially.
 The magnitude of the corner localized eigenstates at zero energy for the finite SSH model are presented in Fig.\ref{fig2}(a,c). The figure depicts the exponential confinement of the boundary states at the corners only despite the presence of degenerate bulk bands. These corner-localized BICs, ideally localized in a zero-dimensional ($0D$) region, are signatures of HOTI\cite{Benalcazar}.
 
There are also zero-energy bulk states in our model (see Fig.\ref{fig2}(b,d)) {that are neither fully extended nor fully localized. Typically they possess localization at the corner sites yet having non-negligible bulk contributions. Hence these states are different from a bulk ZES that does not manifest any corner or edge localizations.} Though weak hopping links from the corners cause the appearance of corner modes or BICs localized at single or multiple corners, {they often hybridize} with the extended or delocalized zero energy bulk modes where the hybridized state gets finite contributions both from the bulk as well as from the localized peak/peaks at the corners. { In Ref.\cite{Benalcazar}, these are termed as corner-localized resonances. Once excited, an electron in such a state remains long lived in the corner only to disappear into the bulk after a long time\cite{Benalcazar}. As opposed to that, an excitation of a corner localized BIC never leaks into the bulk. Only if the protection mechanism for BICs is broken, one can expect to witness these corner-localized resonances in the spectrum.} This selective protection is conjectured to be the selective manifestation of the crystalline and chiral symmetries of the Hamiltonian in the eigenstates\cite{Benalcazar}.

 The existence of corner-localized BICs can be tested directly by adding non-Hermitian on-site terms to the Hamiltonian Eq.(\ref{1}) as performed in Ref.\cite{Benalcazar}. In this context, we should mention that the corner states for a photonic $2D$ SSH model on a square lattice having zero gauge flux and without intra-cellular next-nearest-neighbor (NNN) hopping are not protected topologically and capable to fabrication disorders. However, a new type of corner states arise, after considering the intra-cellular NNN hopping even after the absence of gauge flux, which are robust against certain fabrication noises\cite{xu}. Recently, these gapless corner modes have been observed experimentally for a photonic crystal slab within a parameter space with due consideration of translation as an additional synthetic dimension\cite{dong}.\\
 
\subsubsection{Nonzero energy in-gap states}
 
Fig.\ref{fig1}(c) shows two in-gap bands at nonzero energies ($|E|\approx 2$) in the topological regime and these are in spectral isolation from the bulk bands\cite{Benalcazar}. They start to become visible for $\Delta/t\lesssim -0.5$. The probability distributions of these in-gap modes are reported in the bottom panels of Fig.\ref{fig2}. The figures illustrate that the large amplitudes of these in-gap states are discernible along the one-dimensional ($1D$) horizontal and vertical edge which attenuate exponentially along the direction perpendicular to the edge. Based on the direction of amplitude decay, they are familiar as horizontal ($y$-edge) and vertical ($x$-edge) edge states\cite{dias} (notice Fig.\ref{fig2}(e,f)). We typically call them type-I $1D$ edge states where either $x$-edge or $y$-edge is observed. The existence of the horizontal and vertical edge states can be interpreted assuming the vanishing limit of weak hopping amplitudes. This, in turn, gives rise to decoupled chains at the $x$ or $y$ edges of the lattice developing two edge state bands\cite{dias2}. An in-gap state can also display amplitude decay both in horizontal (for $y$-edge) and vertical (for $x$-edge) directions (notice Fig.\ref{fig2}(g)). We call them type-II $1D$ edge modes, which is a mix of a pair of $x$- and $y$-edge modes.

The edge and corner states are not identical but both of them have topological origins. The former is the result of a two-dimensional Zak phase\cite{liu}, while the latter can be understood from second-order topological effects\cite{edge,dias}.

We may mention here that the bulk states for ribbon configuration feature a similar trend as noticed in the both-side-open square lattice case (discussed above). This is expected because both the configurations have the same bulk lattice\cite{ma}.\\

{\subsection{Polarization and filling anomaly}}

A 2D SSH model preserves time-reversal symmetry and hence does not possess a nonzero Chern number. But having additional $C_4$ symmetry, such system can display filling anomaly and polarization as complementary quantum number to describe its topology\cite{ma}.
  
  In a periodic SSH chain, with bands with $E<0$ completely filled, there remains one electron as well as one positive ion with charge $e$ per unit cell producing charge neutrality in the insulator. But in an system with OBC, scenario depends on the ratio of the hopping amplitudes\cite{benal2}. In the Wannier representation where electrons are described by localized Wannier functions (see Appendix \ref{ap2}), the trivial phase sees the Wannier centers to coincide with the real space unit cell centers - which are also the ionic charge centers. Thus the charge neutrality is preserved. 
  But in the topological phases, the charge neutrality is lost in the sense that positive and negative charge centers do not coincide anymore thereby creating polarizations. The Wannier centers\cite{benal2} in the occupied bands get off-centered from the center of the unit cells resulting in an odd charge imbalance (total charge being odd in units of $e$, the electronic charge) which is distributed equally among two halves of the lattice due to the reflection symmetry. This creates dipole moments and produces filling anomaly such as fractional charge quantizations at the chain boundaries.

In a 2D SSH model, an unit cell is a square and in the topological phase in an system with OBC, a off-centered Wannier function can get localized at the corners and/or at the center of the edges of the unit cells.
Charge neutrality is lost in expense of preserving the $C_4$ symmetry\cite{benal2} and quantized edge and corner charges in multiples of $Q_e=\frac{e}{2}$ and $Q_c=\frac{e}{4}$ respectively are obtained in the topological phase\cite{benal2}.

In short, a reorganization of the number of states within energy-bands, while altering the boundaries from periodic to open, results in the filling anomaly and topological charges in the topological regime\cite{Benalcazar,benal2}. For example, the central band in the system with OBC can show prominent contributions in the density of states from the corner{/edge} unit cells as demonstrated in Ref.\cite{Benalcazar}.  {The lack of homogeneity in both ZES and nonzero energy in-gap states as shown in Fig.\ref{fig2}(a-d) and Fig.\ref{fig2}(e-g) respectively} also support such claim.
\\\\

{\section{Domain Walls in a SSH Lattice}}\label{sec4}
{ As already mentioned, a topological SSH chain hosts boundary modes or end solitons with fractional boundary charges that are quantized by the Zak or Berry phases\cite{mandal}. On introducing
  a topological defect like a domain wall in such a system ($e.g.,$ in Polyacetylene\cite{su,heeger}), which reverses the weak-strong staggered variations in the hopping at the intermediate DW-site, one zero energy domain-wall-soliton localized at the DW-site can be witnessed. But due to breaking of the translational symmetry, a $k-$space integration for Berry phase in such scenario remains no more possible. But one can go to real space Wannier representation\cite{benal2} where Zak phase becomes proportional to the displacement of the Wannier center from the unit cell center at the DW location. In fact, a DW can be treated as an internal boundary within the system and as per the bulk-boundary correspondence, a topological boundary mode localized at the DW can appear in the spectrum.}

Topological defects can appear in $2D$ or $3D$ as well in the form of vortices or magnetic monopoles\cite{jackiw}. The mass-like modulation term $\Delta$ in Eq.(\ref{1}) can take a constant or an spatially inhomogeneous value producing a mass gap or zero energy modes (in the topological phase) in the energy spectrum respectively\cite{jackiw,hou}. Like in Polyacetylene (similar to $1D$ SSH chain with a DW) where defect is created due to Peierls instability, in $2D$ SSH model also such defects can be observed (for example by the Kekul\'{e} distortion in 2D graphene-like structures\cite{hou}). This section aims to study numerically the intricate behavior of ZES and in-gap modes in the presence of static topological defects in our $2D$ SSH model.\\

{\subsection{Intersecting DW lines}}

\begin{figure}[b]
   \vskip .4 in
   \begin{picture}(100,100)
    \put(-220,0){
    \includegraphics[width=.36\linewidth]{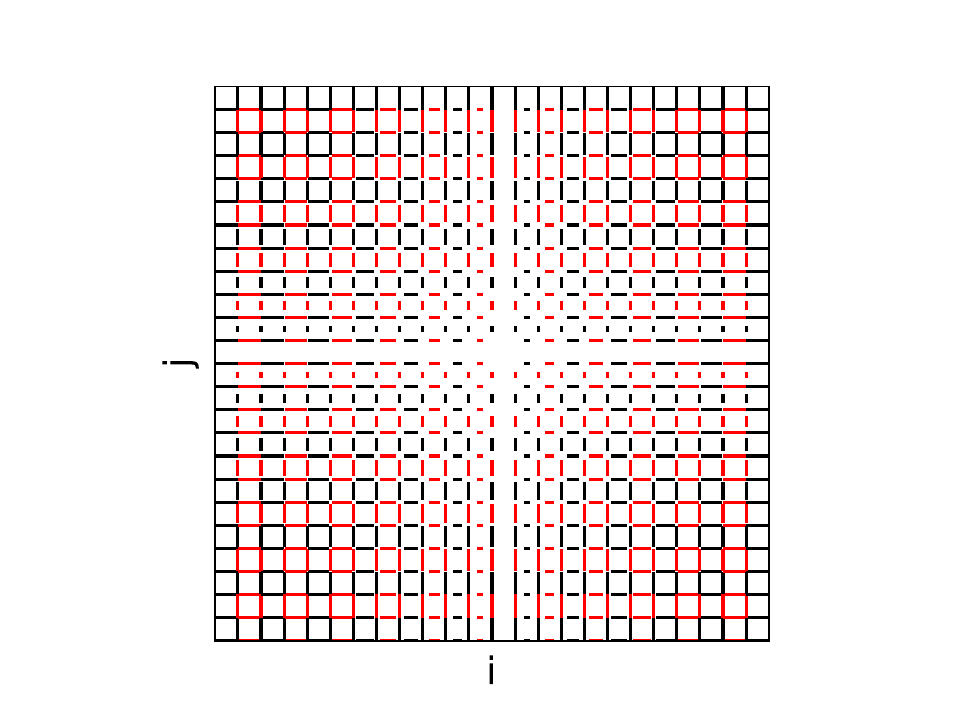}}
    \put(-50,0){
    \includegraphics[width=.32\linewidth]{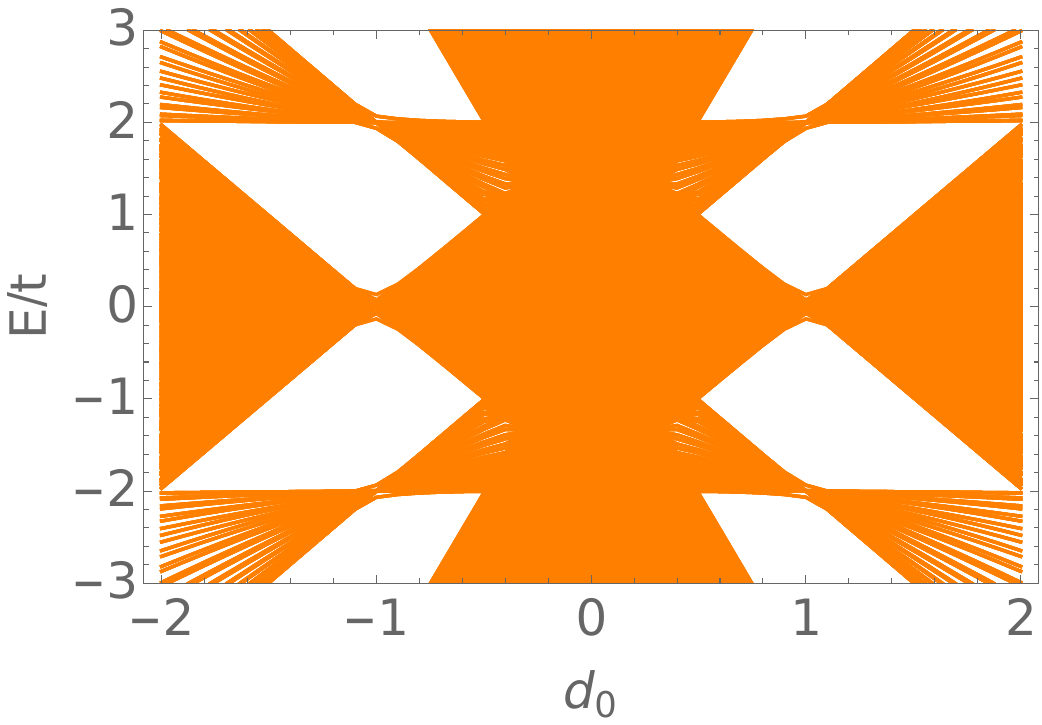}\hskip .2 in
       \includegraphics[width=.3\linewidth]{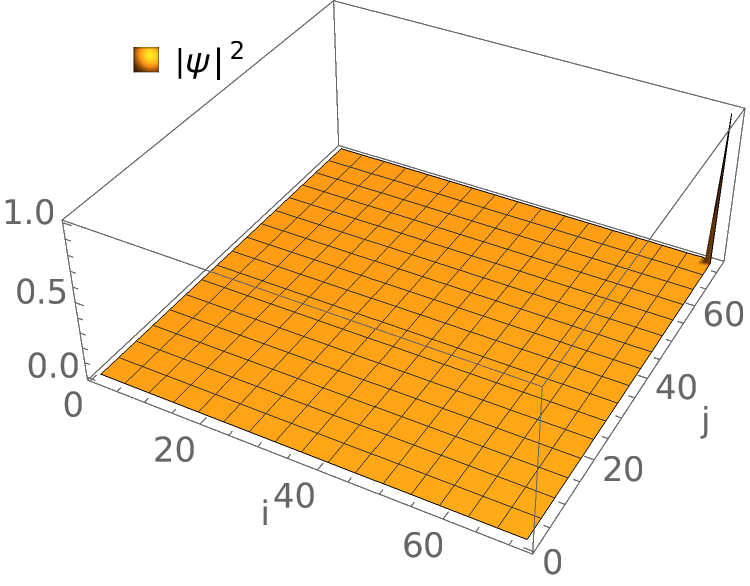}}
    \put(-200,115){(a)}
    \put(40,115){(b)}
       \put(245,115){(c)}
   \end{picture}\\
   \vskip .3 in
   \begin{picture}(100,100)
     \put(-190,0){
       \includegraphics[width=.3\linewidth]{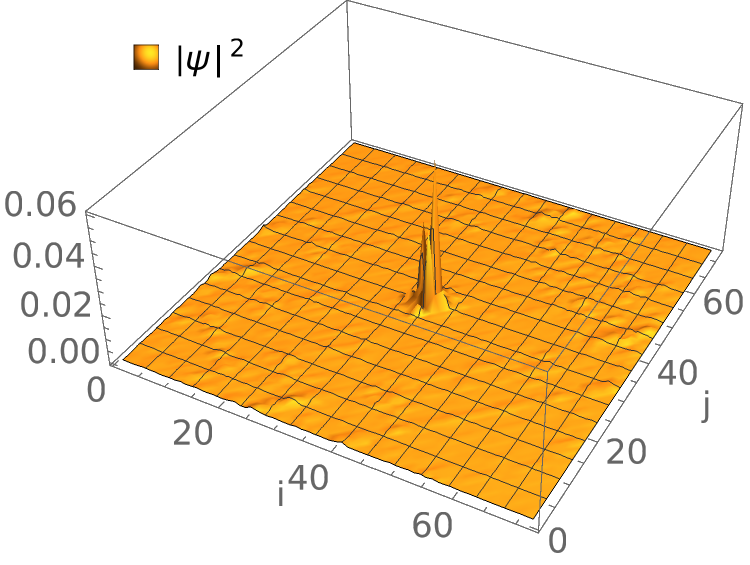}
          \includegraphics[width=.3\linewidth]{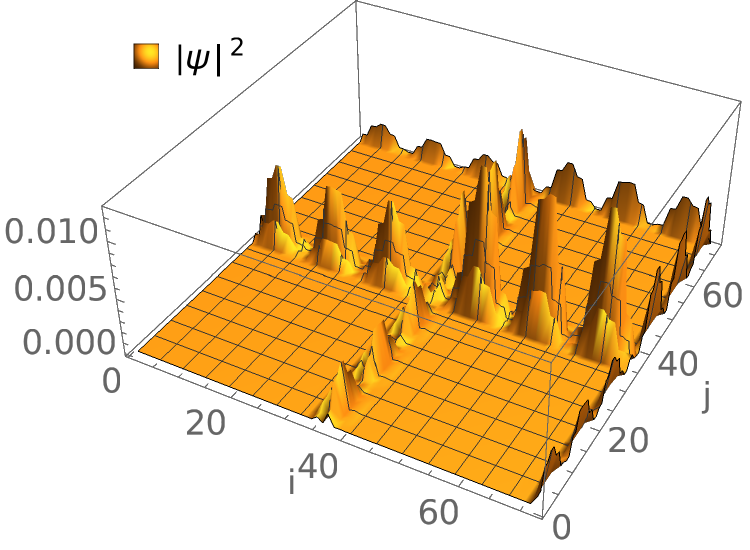}
   \includegraphics[width=.3\linewidth]{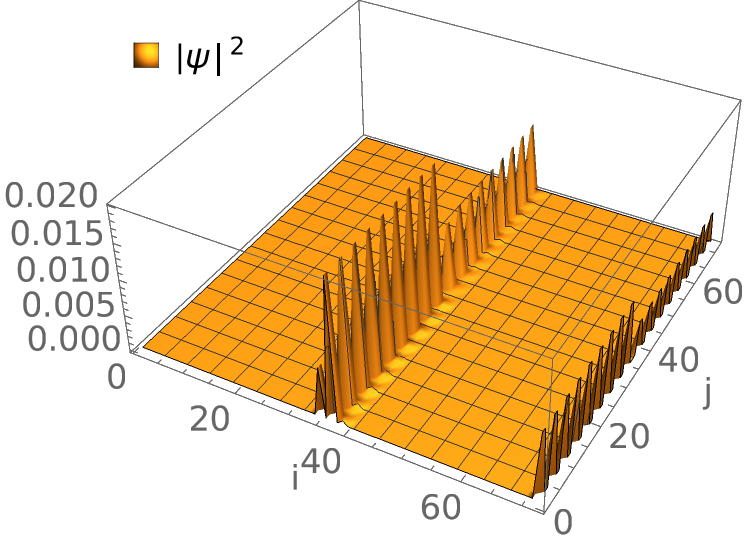}}
   \put(-185,100){(d)}
   \put(50,100){(e)}
     \put(245,100){(f)}
   \end{picture} \\
\caption{(a) Cartoon for distribution of hopping modulations ($\delta_i,\delta_j$) in a $24^2$ lattice with $i_0=j_0=12$ and $\xi/a=8$. Rest of the figures correspond to  $72\times72$ lattices with domain walls at $i_0=j_0=36$ and for $\xi/a=1$. (b) Low energy spectra (with $L$ number of ZES) in a 2D SSH model as a function of domain wall amplitude $d_{0}$. (c,d) show {typical} ZES with corner peaks and DW peaks respectively and (e,f) show two typical in-gap states for $d_0=-0.9$.}
\label{fig6}
\end{figure}

A vortex potential in a honeycomb or Kagome lattice can be given in polar coordinates as $\Delta_{\l}({\bf r})=\Delta(r)e^{il\theta}$ where $l$ denotes the vorticity\cite{hou,chen}. In Ref.\cite{chen}, the vortex potential is considered to be $\Delta({\bf r})=\Delta_{0}\tanh(r/\xi)e^{-i\theta}$ with $\Delta_{0}$ and $\xi$ representing the strength and size of vortex core.
In our case of a 2D square lattice, we first probe the effect of two lines of domain walls in the 2D SSH model (see Fig.\ref{fig6}(a)) which undergo changes in the sign of the hopping modulation at the DW positions given as:
\begin{align}\label{dw1}
  \delta_{i}&=d_0\tanh\Big[\frac{i-i_{0}}{\xi/a}\Big] \cos[(i-1)\theta_x],\nonumber\\
 \delta_{j}&=d_0\tanh\Big[\frac{j-j_{0}}{\xi/a}\Big] \cos[(j-1)\theta_y].
\end{align}
Here $\delta_i$ ($\delta_j$) is the hopping modulation for hopping along $x~(y)$ at $x=ia~(y=ja)$ respectively. $d_0$, $\xi$ and $a$ represent the maximum strength of the modulation, the width of the domain wall (DW) and the lattice constant respectively. So it indicates two perpendicular lines of DWs at $x=i_0a$ and $y=j_0a$ that intersect at $(x,y)=(i_{0},j_0)a$ which we can call the center of DWs, as demonstrated for $\theta=\pi$ by the cartoon in Fig.\ref{fig6}(a) where different colors (length) of bonds represent different signs (strengths) of modulations $\delta_i,\delta_j$'s.
In the present study, we typically consider $i_0=j_0=L/2$, though results for other DW positions are reported as well.

\begin{figure}
   \vskip .15 in
   \begin{picture}(100,100)
     \put(-190,0){
  \includegraphics[width=.3\linewidth]{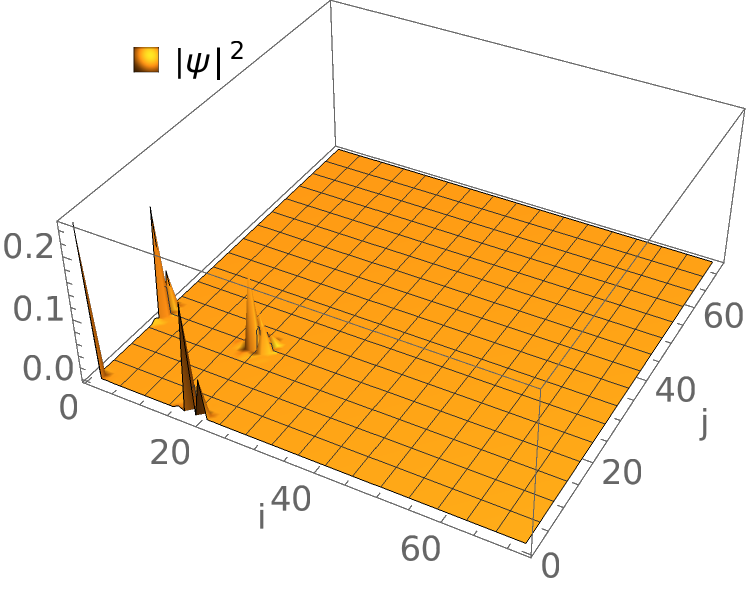}
  \includegraphics[width=.3\linewidth]{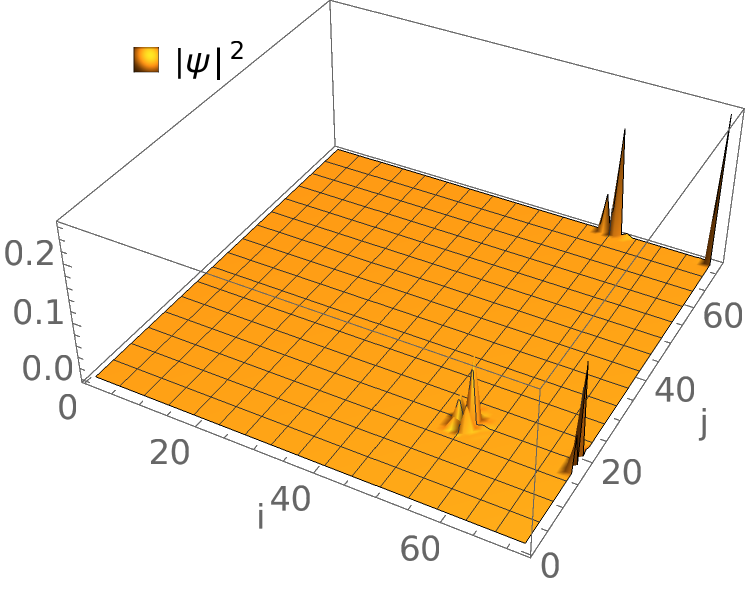}
     \includegraphics[width=.3\linewidth]{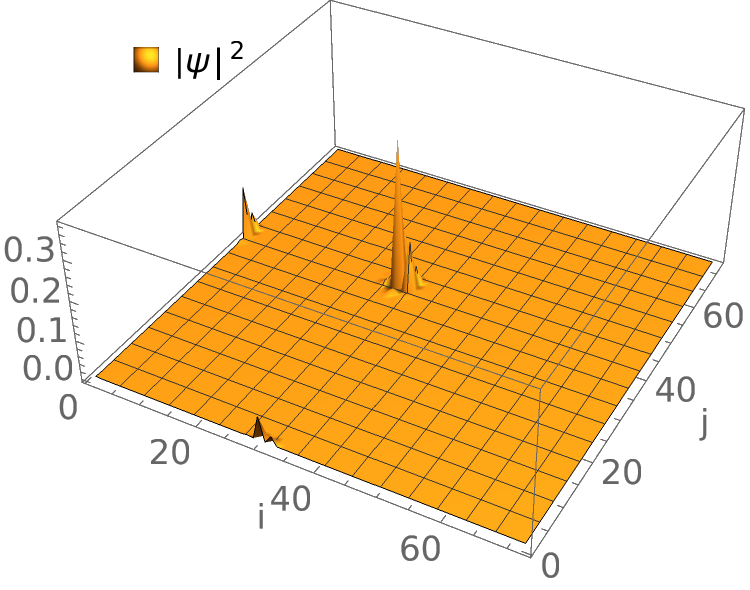}}
     \put(-180,110){(a)}
     \put(-20,110){(b)}
        \put(135,110){(c)}
   \end{picture}\\
   \vskip .25 in
   \begin{picture}(100,100)
     \put(-130,10){
  \includegraphics[width=.30\linewidth]{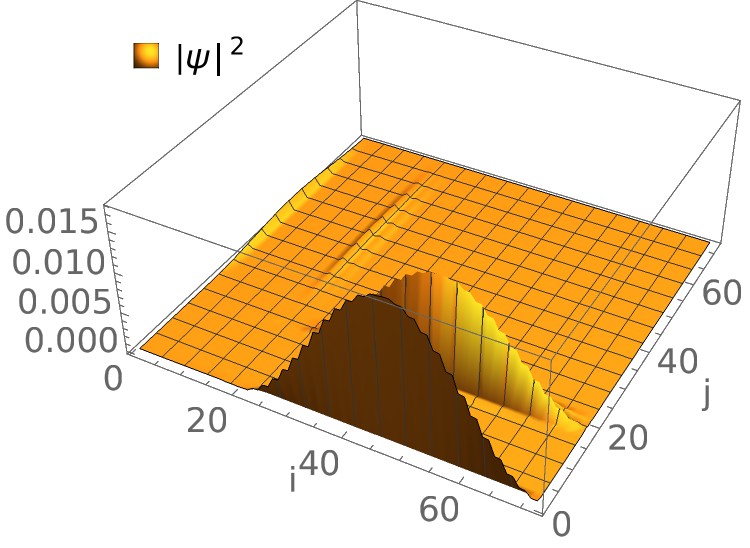}~~~~
     \includegraphics[width=.30\linewidth]{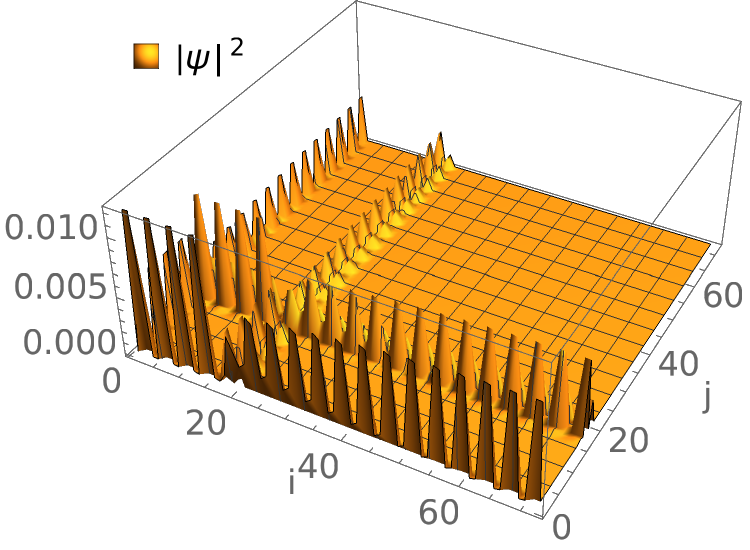}}
     \put(-120,110){(d)}
        \put(50,110){(e)}
   \end{picture}
\caption{Density plots for $|\psi|^2$ corresponding to ZES localized at DW for DW position (a) $i_0=j_0=18$, (b) $i_0=54,~j_0=18$, (c) $i_0=30,~j_0=42$ on a $72\times72$ lattice. (d,e) show two typical nonzero energy in-gap states with  $i_0=j_0=18$. We consider $d_{0}=0.9$ and $\xi/a=1$ in all the plots.}
\label{fig7}
\end{figure}

What follows here is a discussion on having such set of domain walls for $\theta=\pi$. The energy spectra as a function of domain wall amplitude $d_{0}$ is presented in Fig.\ref{fig6}{(b)}. Like in 1D, here in 2D square lattice also the inclusion of DWs result in additional in-gap states {(often called bound states\cite{domain})}. The localized ZES show localization either at one corner or at the DW center(see Fig.\ref{fig6}(c,d)).
We have seen that the DW of the 1D SSH chain connects two
dimerized phases {($i.e.,$ each site has different hopping amplitudes with its two neighbors whose order flips at the DW)} of the chain where the absence of dimerization at the DW, the so-called unpaired site, produces {a zero mode, famously called a Jackiw–Rebbi mode}\cite{rebbi}. {In 2D, sign flips occur in $\delta_i$ and $\delta_j$ at the DW locations (see Eq.~\ref{dw1}). Hence for the site ($i_0,j_0$) where the orthogonal DW lines cross, there remains no dimerization with any neighbor whatsoever ($i.e.,$ all four neighbors are connected via same hopping amplitudes). In the bipartite language this creates a sublattice
imbalance locally and more generally, a topological defect that produces a new zero-energy state.}
Interestingly, for the modes localized at DW positions, we also witness some wavefunction peaks at the $x$ (or $y$) edges with $y$ (or $x$) coordinate same as that of the DW center (see Fig.\ref{fig7}(a-c)).  The number of zero energy modes do not change (compared to the topological regime of the 2D SSH model) by the insertion of the DW, as long as we set $i_0=j_0$
but it reduces otherwise.  Like in 1D, the zero modes disappear for $d_{0}=0$ where one gets a nearest neighbor (NN) tight binding model with no zero energy state in a finite lattice. The ZES vanishes for very large $d_0$ as well.
The in-gap states localize not only along the edges but also along the perpendicular DW lines as shown in Fig.\ref{fig6}(e-f).
This is because the perpendicular DW lines act as internal 1D interfaces where the dimerization changes
sign, just like at physical edges. Mathematically, the
DW line is a locus where the bulk mass term changes
sign and a Jackiw–Rebbi domain-wall mode\cite{rebbi} is confined to
that line. Thus, in addition to edge localized states, the system also features
 states localized along the DW lines.


Typical probability density plots for ZES and nonzero energy in-gap states with DW center at and off the center of the lattice are shown in Fig.\ref{fig6} and Fig.\ref{fig7} respectively. Like in the cases without DWs, here also the spectrum features a continuum of ZES containing a number of topological as well as bulk states. But unlike the previous case, localized ZES are also obtained at the DW position or at $x$ or $y$ positions same as that of the DW center. For the nonzero energy in-gap states, we not only see finite wavefunction amplitudes at the edges but also along the DW positions.
Notice that for $d_0<0$, the hopping amplitudes around the crossing point $(i_0,j_0)$ becomes smaller preferring an isolated peak in the DW solitonic state whereas for $d_0>0$, the opposite scenario occur and the DW soliton state remains less localized at $(i_0,j_0)$ with additional localizations at the boundaries corresponding to negative $\delta$, as can be seen from the ZESs in Fig.\ref{fig6} and Fig.\ref{fig7}. \\

{\subsection{Loop of DWs}}

Distribution of the DWs {outlined in Eq.\ref{dw1}}, however, can't produce vortex-like structures in the wavefunction as they are not radially symmetric. So we consider a radial distribution of DWs given by
\begin{align}\label{dw2}
  \delta_{i}&=d_0[1-2e^{-\frac{|r-r_{0}|}{\xi/a}}] \cos[(i-1)\theta_x],\nonumber\\
 \delta_{j}&=d_0 [1-2e^{-\frac{|r-r_{0}|}{\xi/a}}]\cos[(j-1)\theta_y].
\end{align}
\begin{figure}
  \vskip .15 in
   \begin{picture}(100,100)
     \put(-190,0){
  \includegraphics[width=.36\linewidth]{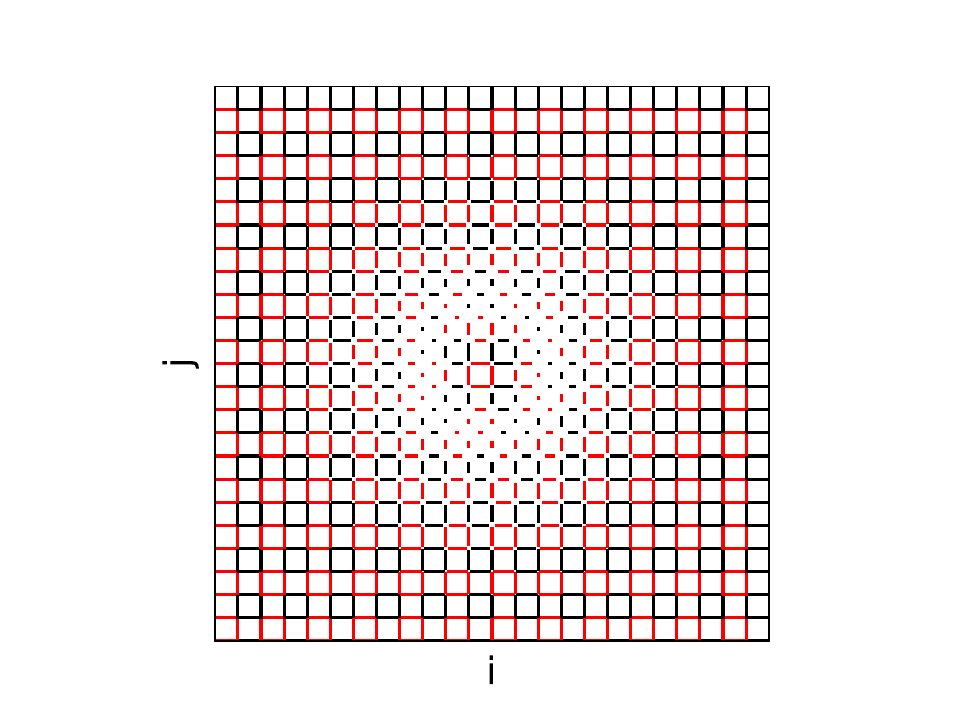}
  \includegraphics[width=.3\linewidth]{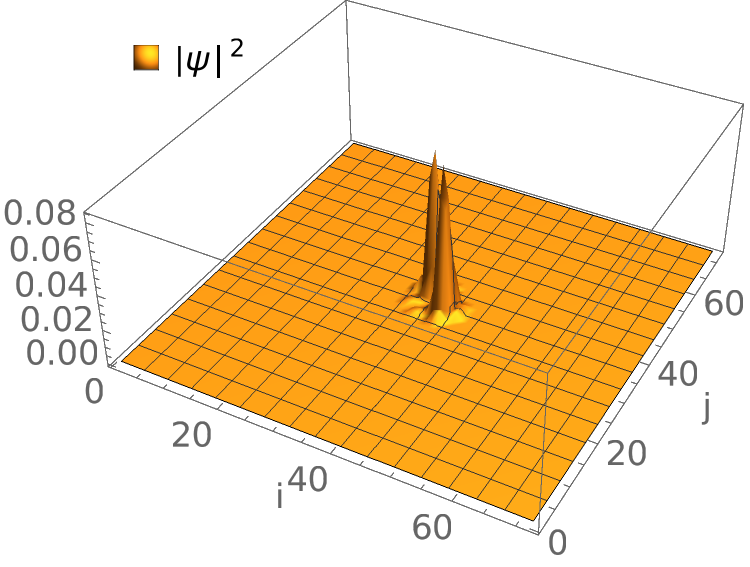}
     \includegraphics[width=.3\linewidth]{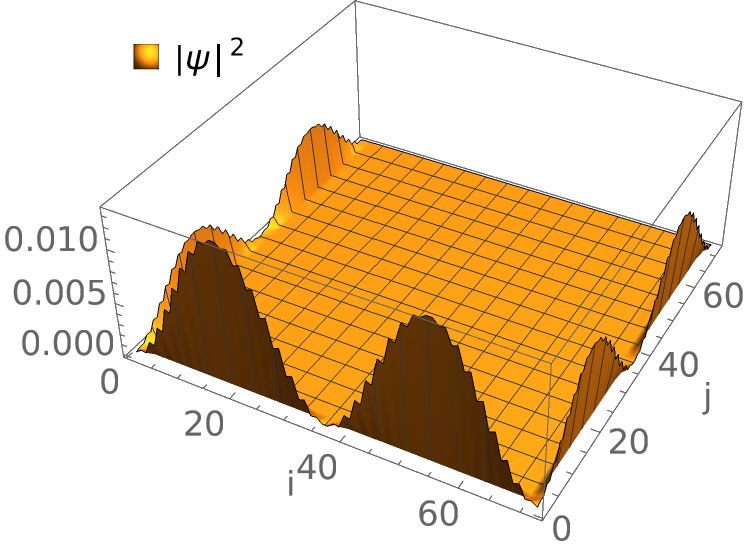}}
      \put(-170,110){(a)}
     \put(110,110){(b)}
        \put(260,110){(c)}
   \end{picture}\\
   \vskip .25 in
   \begin{picture}(100,100)
     \put(-180,0){
  \includegraphics[width=.3\linewidth]{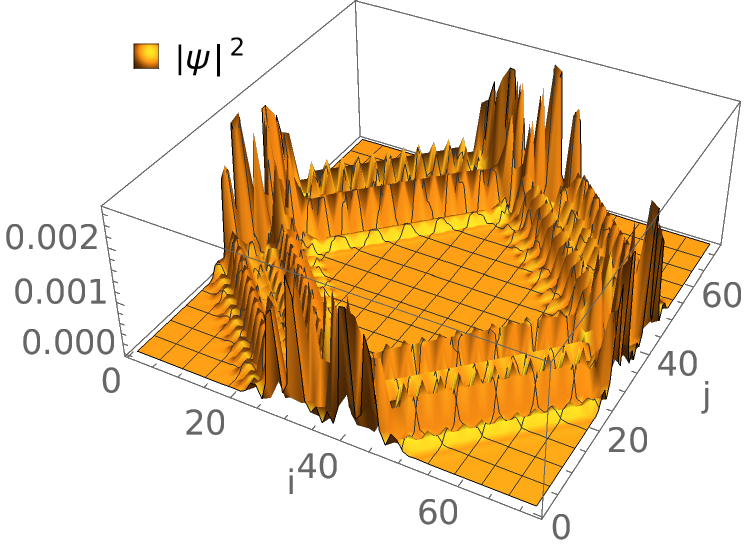}
  \includegraphics[width=.3\linewidth]{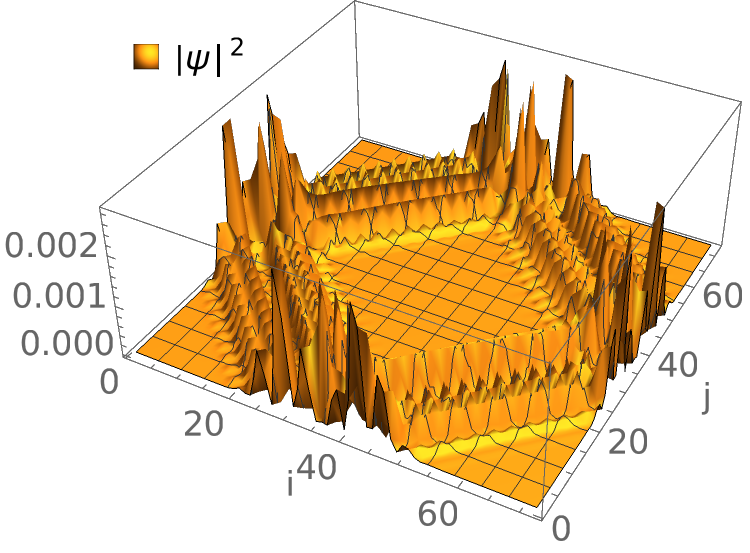}
     \includegraphics[width=.3\linewidth]{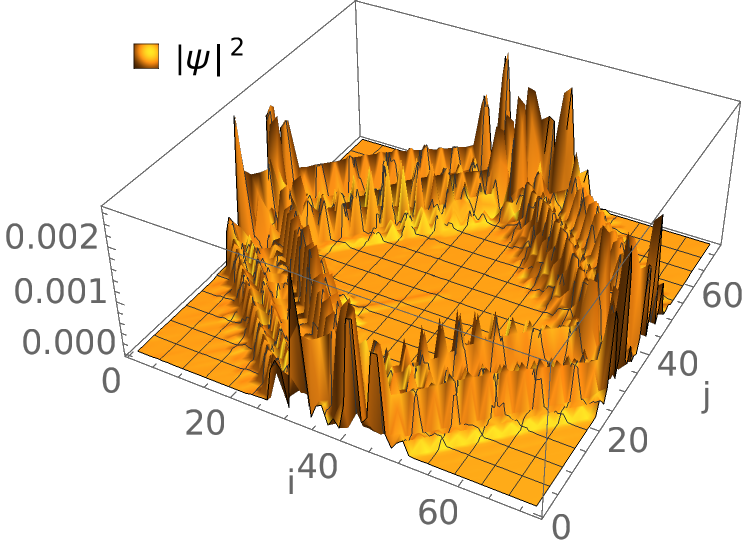}}
     \put(-50,100){(d)}
     \put(110,100){(e)}
        \put(260,100){(f)}
   \end{picture}
   
\caption{(a) Cartoon for distribution of hopping modulations ($\delta_i,\delta_j$) in a $24^2$ lattice with $r_0=(\frac{L-1}{2},\frac{L-1}{2})a$ and $\xi/a=4$. Density plots for $|\psi|^2$ corresponding to two typical in-gap states (b,c) and a ZES (d) for $\xi/a=2$ on a $72\times72$ lattice for $d_{0}=1.2$. (d,e) show typical ZES for (e) $\xi/a=4$ and (f) $\xi/a=6$. }
\label{fig}
\end{figure}
where $r_0$ and $r$ represent the radial coordinate for the DW center and that of the point $(i,j)$ respectively. 
Fig.\ref{fig}(a) gives a cartoon for the distribution of hopping modulations in a $24\times24$ lattice for $\xi/a=4$ where the strength of a $\delta$ is proportional to the length of the hopping links in it while different signs of it correspond to different colors. One can readily identify the circular ring with $\delta=0$ around the location $(i_0,j_0)$. But even though the modulation suggests a circular DW, the Hamiltonian is still lattice-based and thus results in $\xi$ dependent square-like {amplitude} distribution in the ZES (see Fig.\ref{fig}(d-f)).
Notice that with increase in $\xi$, the region with finite wave-amplitudes spread more within the lattice (see Fig.\ref{fig}(d-f)).
Interestingly, the in-gap states (in Fig.\ref{fig}(b,c)) remain localized at the $(i_0,j_0)$ position {or along the edges.}

{ The characteristics of all the localized modes in presence of DWs for both the types that we discuss here, can be understood following the features of a SSH chain with a DW where one ZES gets localized at the DW apart from a second one localized at one physical edge\cite{mandal}. A DW is like an internal boundary in a system where localized ZES can appear as an edge mode in such a boundary.  A 2D set-up with hopping modulations given by Eq.~\ref{dw1} corresponds to assemblies of parallel SSH chains both along $\hat x$ and $\hat y$ directions with DWs at $x=i_0a$ and $y=j_0a$ respectively. Thus all these chains prefer localizations of ZES along the lines: $i=i_0$ and $j=j_0$. However, such linear localizations get disrupted due to two SSH chains along those lines unless we are at the DW center at $(i_0,j_0)$. At this point, localizations due to two chains get reinforced and survive to give states as shown in Fig.\ref{fig6}(d). Similarly, the corner localizations can also survive in some ZES. The nonzero energy in-gap states however show localization along the DW lines as well as the physical boundary. In the second case of loop of DWs with hopping modulations defined by Eq.~\ref{dw2}, the DWs orient along a loop within the square lattice and ZES localizations occur at those DW positions. Corner localizations can also be seen in addition (not shown in Fig.\ref{fig}), as in the previous case.
  }
\\

\section{Quasiperiodic Disorders}
The chiral symmetry of the SSH Hamiltonian is instantly broken via a diagonal disorder. Now out of various types of diagonal disorders, here we consider a quasiperiodic disorder for its realization in many quasiperiodic systems such as Hofstadter butterfly or quasicrystalline crystals\cite{quasicrystal} and the rich localization behavior it shows based on its strength. More specifically, we attempt to explore the outcome of quasiperiodic Aubry-Andre (AA) potential\cite{aubry} in our 2D lattice. There can be different ways to implement it. {Here we discuss two types of quasiperiodic disorders as mentioned below.}

{\subsection{Disorder: Type I}}

 Following Ref.\cite{prb101-014205}, we first choose the disordered Hamiltonian to be 
\begin{eqnarray}
  H&=&H_{SSH}+H_{disorder}\nonumber\\
  &=&H_{SSH}+\lambda\sum_{i,j}^L\cos(2\pi\beta i)\cos(2\pi\beta j)c_{i,j}^\dagger c_{i,j}
  \label{aa1}
\end{eqnarray}
where $H_{SSH}$ represents the SSH type Hamiltonian that we discussed so far, $\beta=\frac{\sqrt{5}-1}{2}$, an irrational Diophantine number and $\lambda$ is the disorder strength.
\begin{figure}
  \begin{center}
   \vskip .01 in
   \begin{picture}(100,100)
    \put(-194,0){
      \includegraphics[width=.4\linewidth,height=1.5 in]{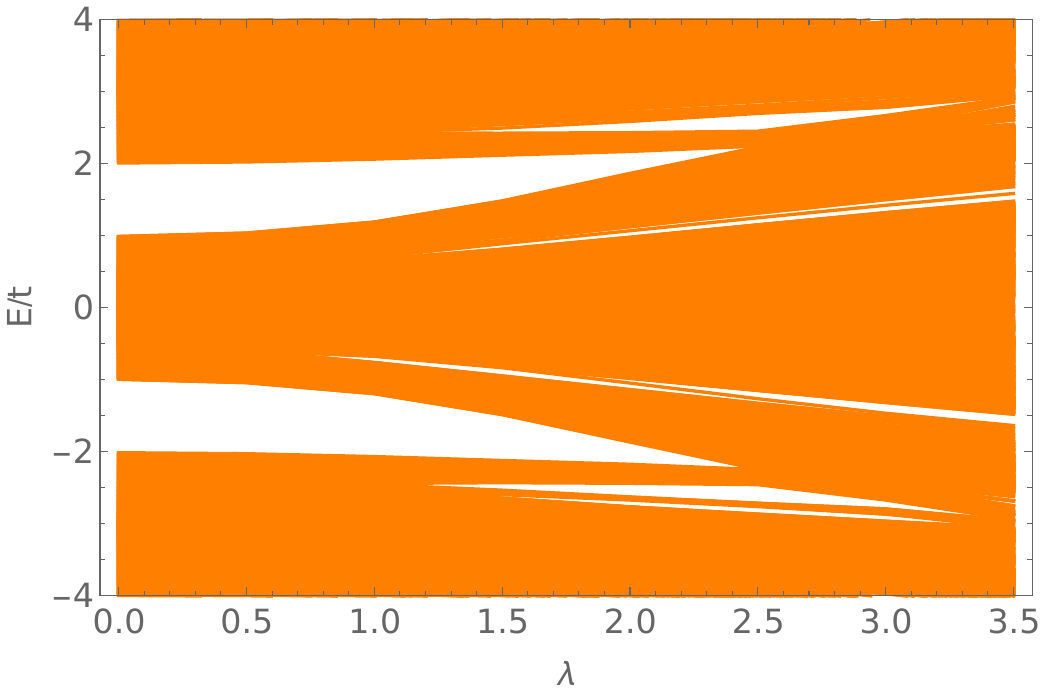}~~~~~
      \includegraphics[width=.5\linewidth,height=1.5 in]{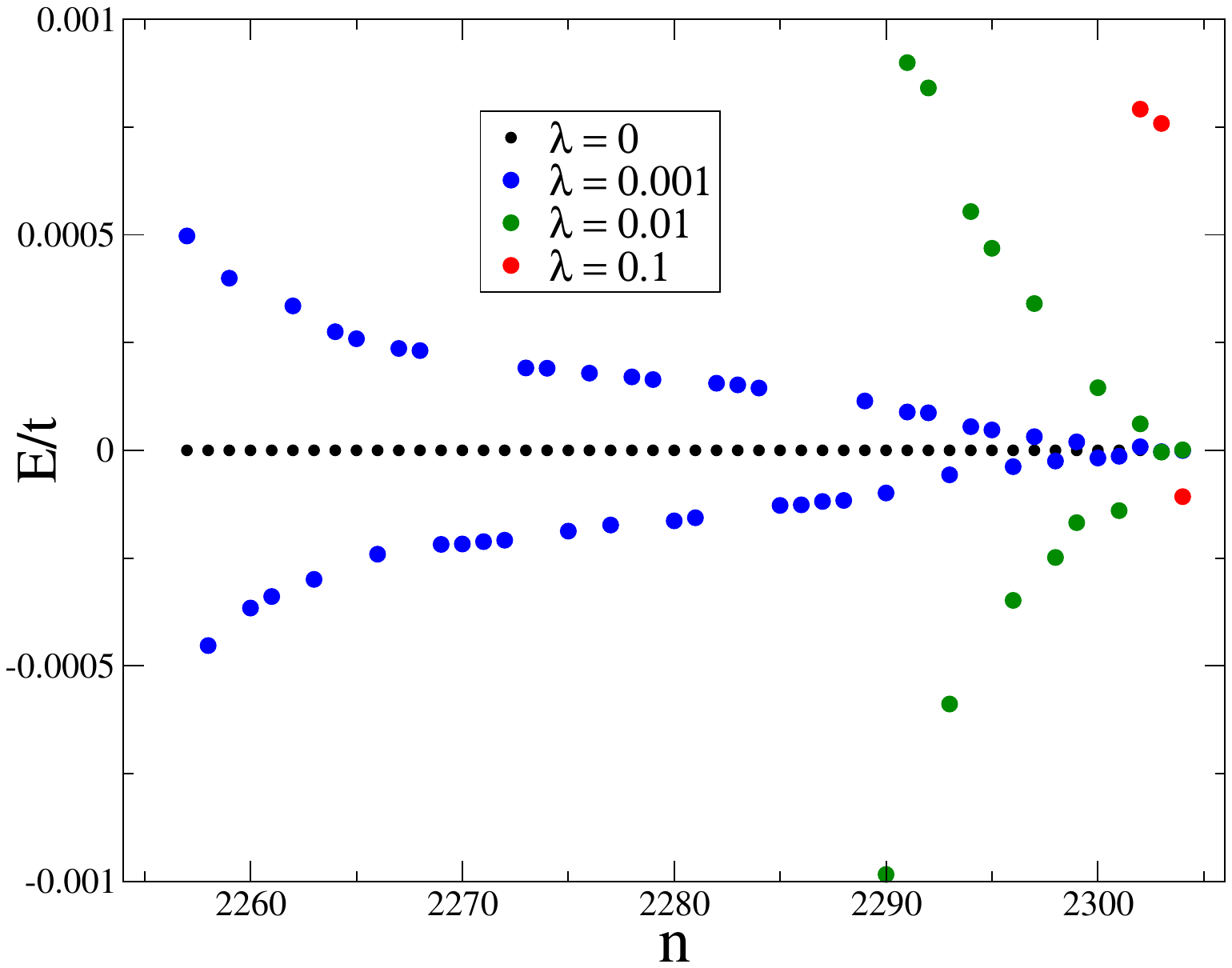}}
    \put(-130,85){(a)}
    \put(100,85){(b)}
   \end{picture}\\
   \vskip .5 in
   \begin{picture}(100,100)
     \put(-184,0){
   \includegraphics[width=.3\linewidth]{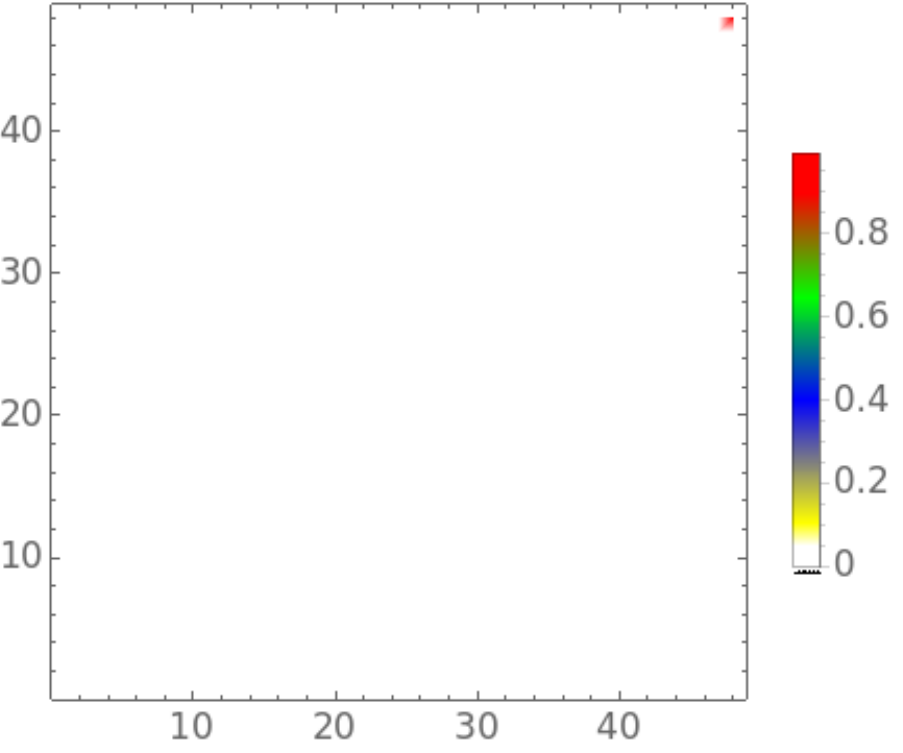}
   \includegraphics[width=.3\linewidth]{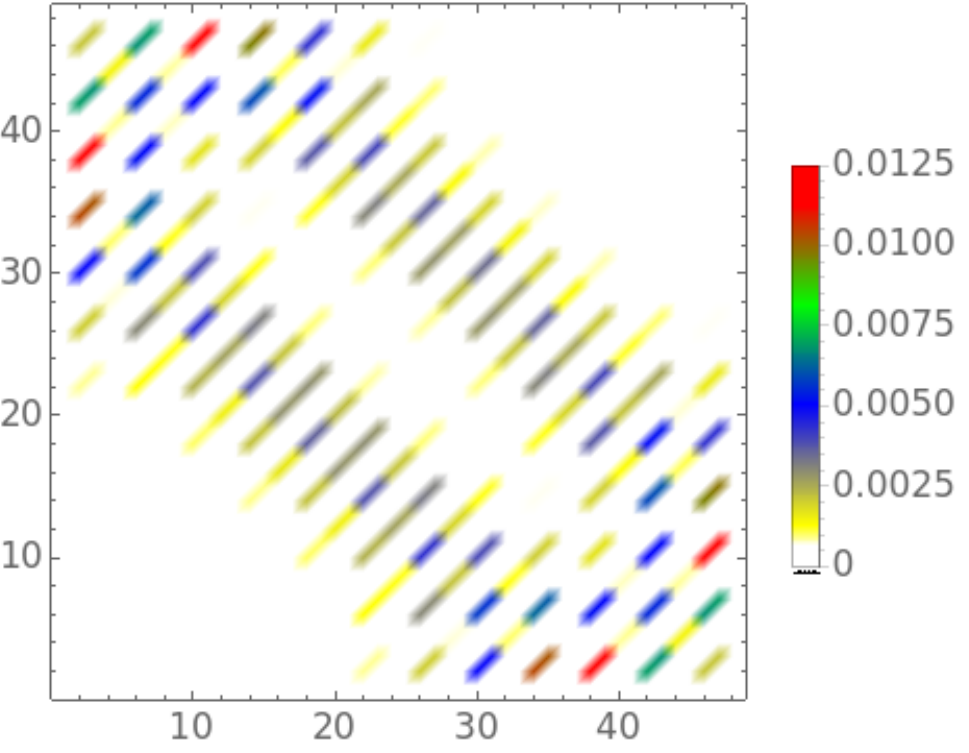}
   \includegraphics[width=.3\linewidth]{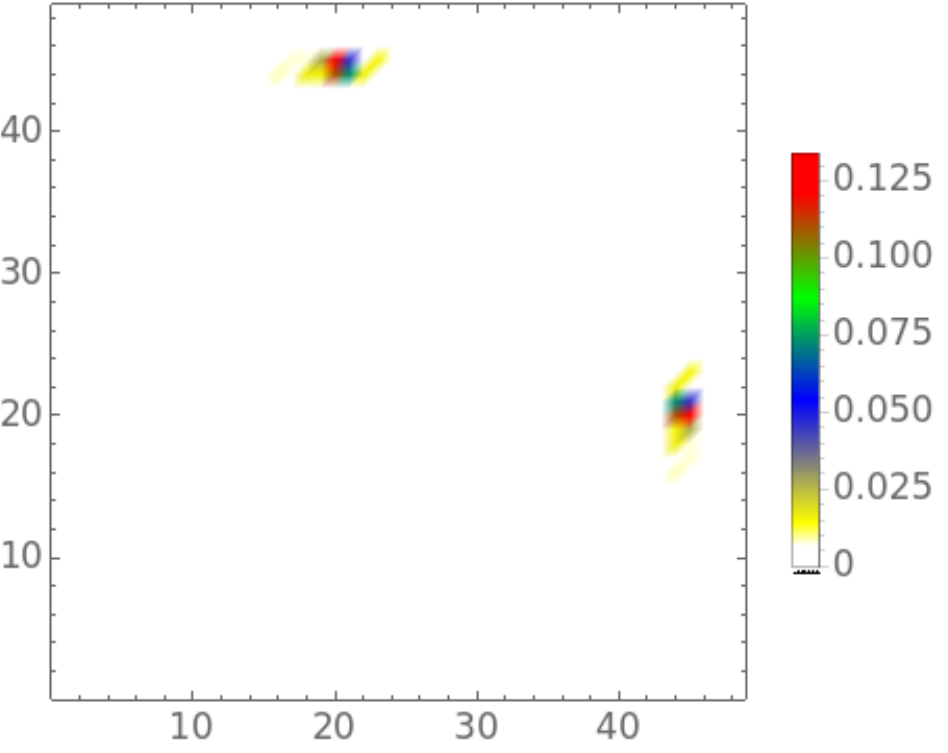}}
       \put(-130,100){(c)}
\put(60,100){(d)}
     \put(210,100){(e)}
   \end{picture}  
   \vskip 0.5 in
   \begin{picture}(100,100)
     \put(-184,0){
   \includegraphics[width=.3\linewidth]{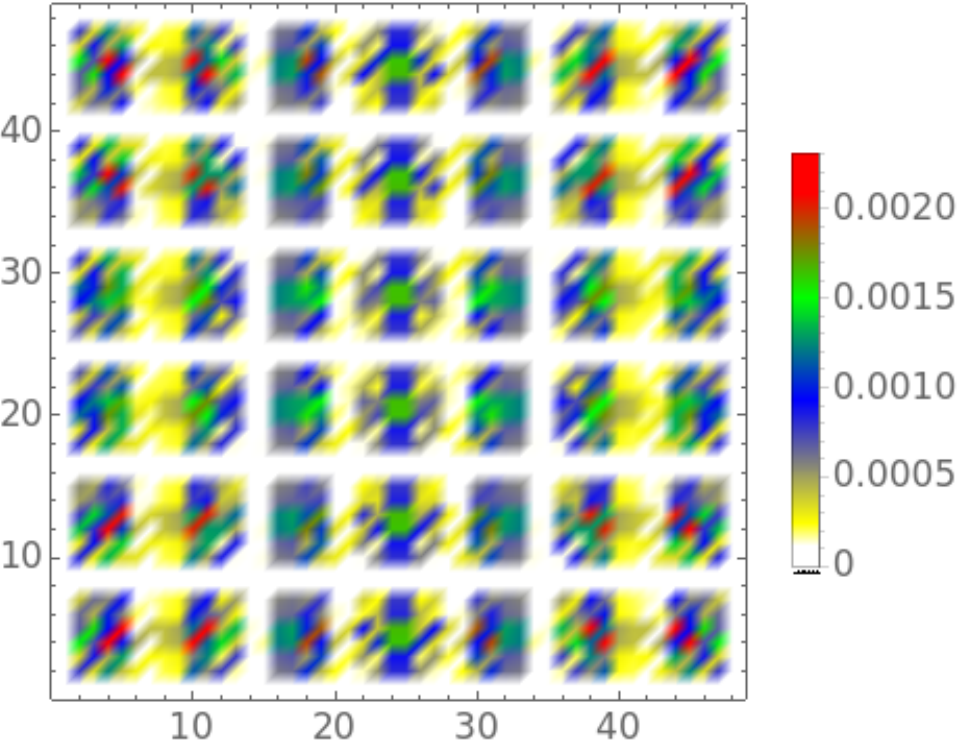}
   \includegraphics[width=.3\linewidth]{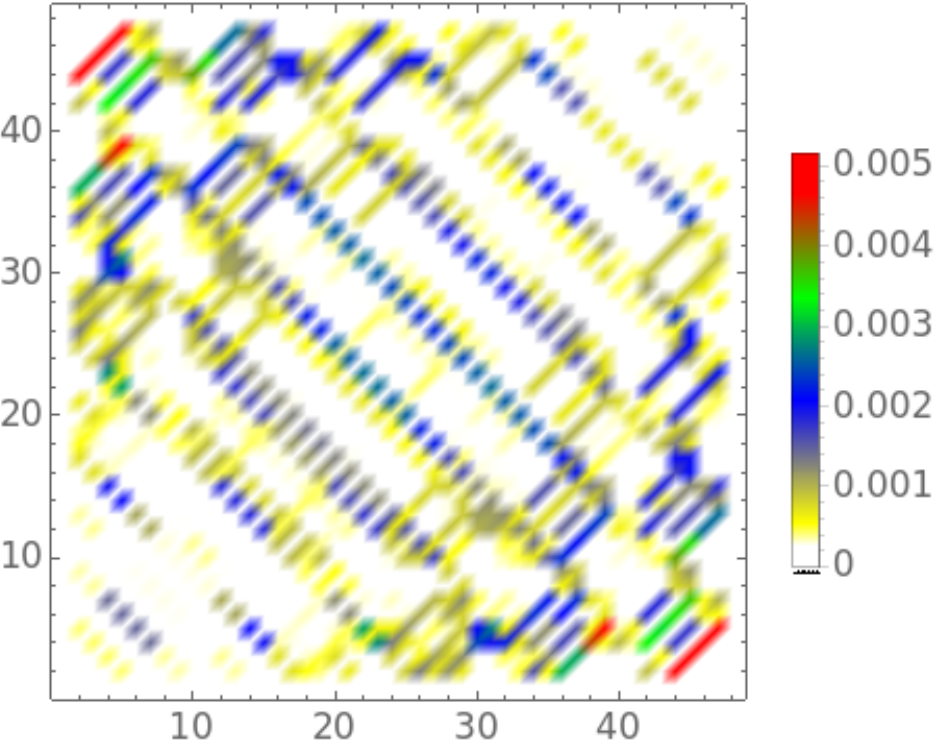}
   \includegraphics[width=.3\linewidth]{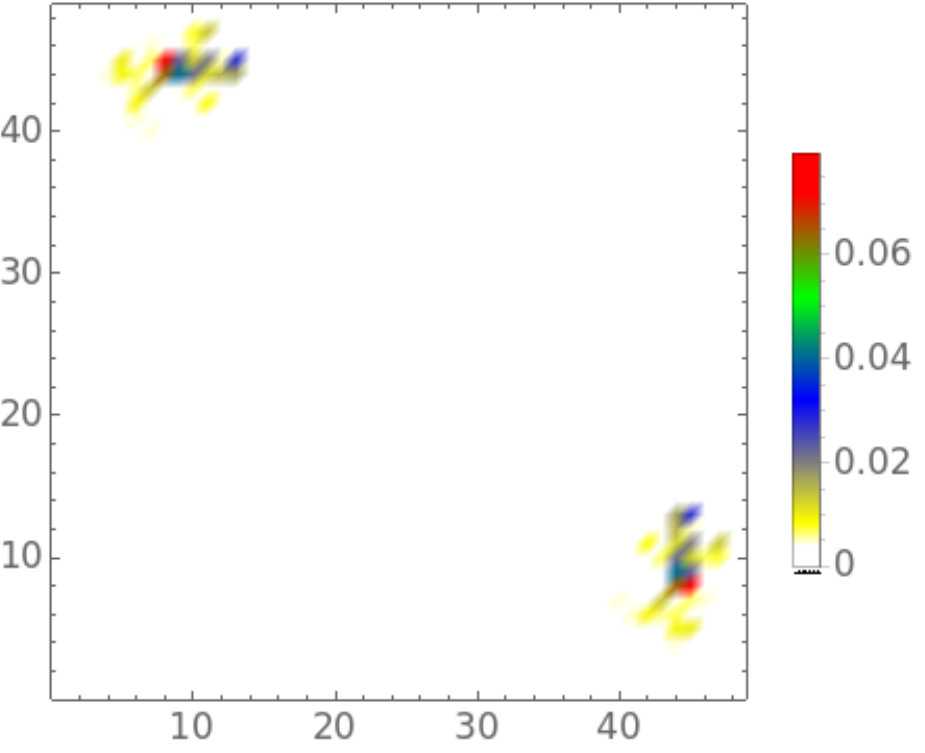}}
   \put(-130,120){(f)}
   \put(60,120){(g)}
     \put(210,120){(h)}
   \end{picture}
\caption{(a) Energy spectra for $\Delta=0.5$ ((b) low energy spectra for various small $\lambda$'s for $\Delta=-1.2$) for uniform disorder in a $48\times48$ lattice. (c)-(h) show $|\psi|^2$ at $\Delta=-1.2$ for typical topological corner states (c-e) and bulk states (f-h) at $\lambda=0$ (c,f), 0.01 (d,g) and 2.5 (e,h) respectively.}
\label{fig10}
\end{center}
\end{figure}
\begin{figure}[htbp]
  \hspace {2in}
  \includegraphics[width=.8\linewidth]{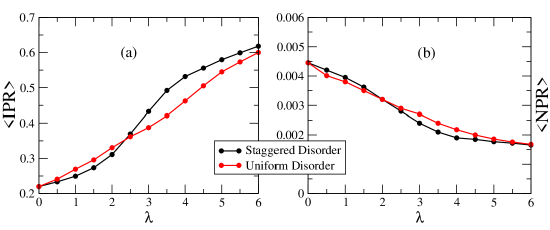}\\
  \includegraphics[width=.6\linewidth]{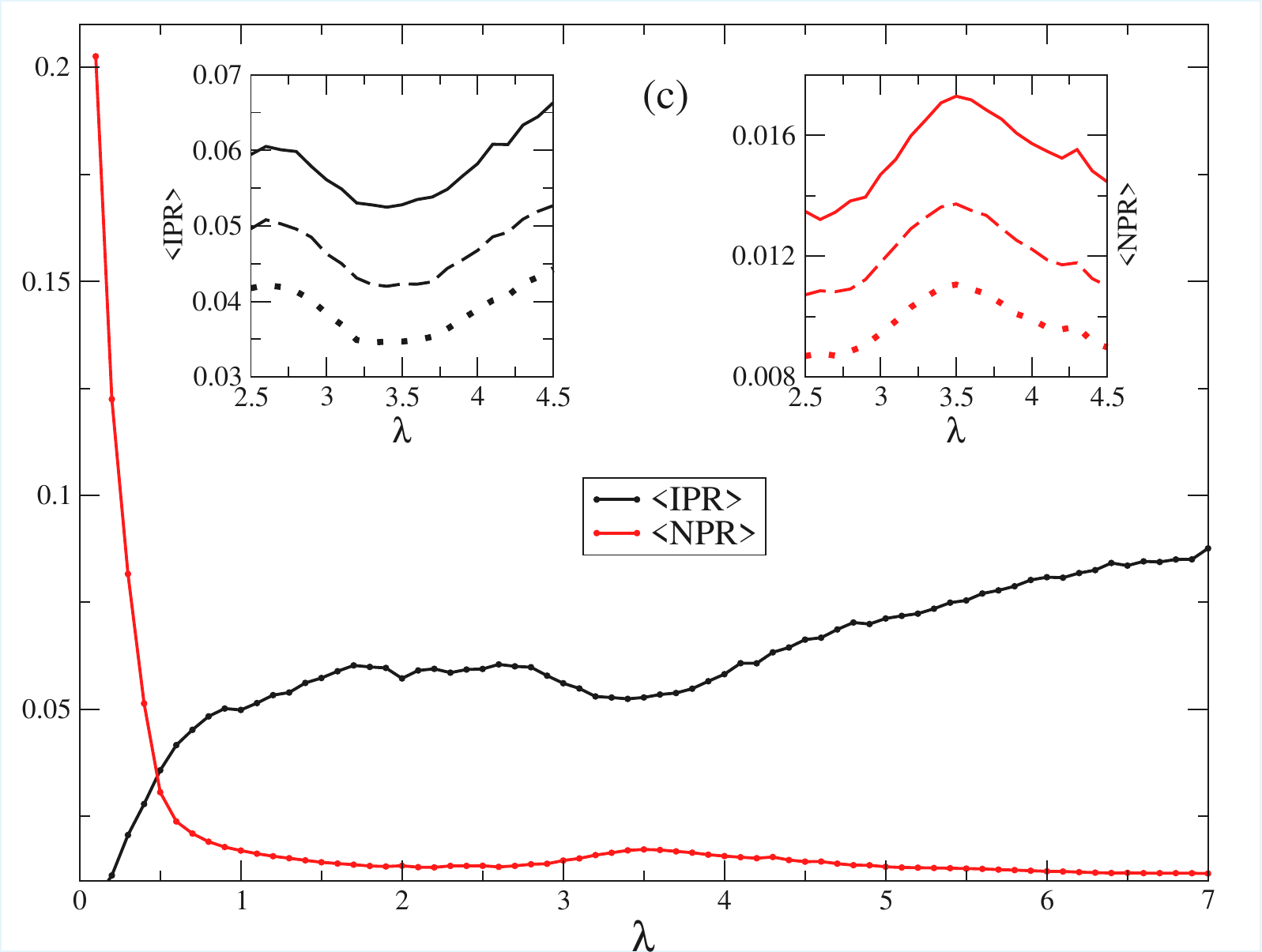}
  \vskip -.1 in
  \caption{[Color online] (a) and (b) show respectively the $<IPR>$ and $<NPR>$ variations respectively with respect to $\lambda$ in a $32\times32$ lattice for uniform and staggered disorders, as per Eq.\ref{aa1}. (c) gives $<IPR>$ (black solid line) and $<NPR>$ (red/gray solid line) values in a $48\times48$ lattice for the staggered disorder given by Eq.\ref{aa2}. The insets there display zoomed-in values for $<IPR>$ (left) and $<NPR>$ (right) that include also results from $64\times64$ (dashed) and $80\times80$ (dotted) lattices.}
  \label{fig18}
  \end{figure}

Such a quasiperiodic potential in a nearest neighbor tight binding model can cause a localization transition in 2D\cite{prb101-014205} by which states get localized beyond a critical $\lambda$. For an SSH model with staggered hopping modulations, such disorder immediately breaks the chirality (see Fig.\ref{fig10}(a)) and it shows no ZES with the low energy spectrum moving away from the $E=0$ axis as the disorder strength is gradually tuned up from zero (see Fig.\ref{fig10}(b)). Within the critical disorder strength corresponding to the localization transition, we see neither fully localized nor fully extended eigenstates but rather partially extended states are witnessed in the system. Evolution of typical corner and bulk states with $\lambda$ are shown in Fig.\ref{fig10}(c-h)). The localized topological states first get delocalized due to disorder which then again gets localized beyond a higher critical $\lambda$. In order to study the interplay between hopping dimerization and disorder, one can also use a staggered $\lambda$ (by replacing $\lambda$ with $\lambda*(-1)^{i+j}$ in Eq.\ref{aa1}) instead of an uniform one and compute the inverse participation ratio (IPR) and normalized participation ratio (NPR) for the eigenstates to witness its localization behavior. An IPR {and NPR for the $n$-th eigenstate are defined, respectively} as
\begin{equation}
  IPR_n=\sum_{i=1}^{L^2}|\phi_n^i|^4,\  NPR_n=(L^2\sum_{i=1}^{L^2}|\phi_n^i|^4)^{-1},
\end{equation}
$\phi_n^i$ being the $n$-the eigenfunction at the site $i$.
It estimates the degree of localization in an eigenstate designated by subscript $n$.
The {overall nature of localization in the system, however, is better understood from the average of these quantities.} In Ref.\cite{s-basu}, an average $<IPR>$ and $<NPR>$ has been calculated in a 1D disordered SSH chain and they find out a nontrivial reentrant localization transition for staggered disorder strengths. When we perform the same in our 2D system following Eq. \ref{aa1}, we find steady behavior of $<IPR>$ and $<NPR>$ with $\lambda$, as shown in Fig.\ref{fig18}(a,b) for $\Delta/t=0.5$, with no re-entrance of localization as seen in 1D.

{\subsection{Disorder: Type II}}

{In search of the exotic re-entrant behavior, we adopt a different distribution of disorder which shows signs of such reentrant localizations}. For example, with the following disordered term
\begin{equation}
  H_{disorder}=\lambda\sum_{i}^L\sum_{j}^L(-1)^{i+j}\cos[2\pi\beta(i+j)]c_{i,j}^\dagger c_{i,j}
  \label{aa2}
\end{equation}
we witness reentrant behavior as shown in Fig.\ref{fig18}(c) for $\Delta/t=-0.8$ in a $48\times48$ lattice. Though at the first localization transition at $\lambda\sim2$, $<NPR>$ does not vanish completely, to some good approximation we can call the unorthodox behavior immediately after that first transition to be a reentrant behavior.  The insets in Fig.\ref{fig18}(c) highlight the exact $\lambda$ window where such nontrivial behavior shows up, also for larger lattices {(upto $64\times64$ square lattices, the limit till which our numerical resources permit us to study)}.
As one can see, the reentrant phase exists roughly between $2.5<\lambda<5.0$. Within this region we witness simultaneous presence of $<IPR>$ and $<NPR>$ and hence indicating a second window for partially localized/extended states there. {A finite size analysis can confirm the existance of this non-monotonic behavior even for large lattices. In fact, we use the peak values of $<NPR>$ at $\lambda\sim3.5$ (see Fig.\ref{fig18}(c) inset) for different lattice sizes to predict its value at the thermodynamic limit via a linear extrapolation on a $<NPR>$ vs. $L^{-2}$ plot and we find it to be non-negligible ($<NPR>~\sim0.008$).} Notice that such coexistence of finite $<IPR>$ and $<NPR>$ also occurred below $\lambda=2$ and adjacent to it. Hence, this system indicates the existence of two single particle mobility edges (SPME) that separate localized and extended states, as also found in 1D\cite{s-basu}.

Usually reentrant behavior emerges where two periodic or quasiperiodic scales clash, like the on-site staggered quasiperiodic potential competing with periodic hopping modulations in our present case. Unlike the uniform quasiperiodic potential where a self-duality symmetry\cite{aubry} causes all states in the spectrum to localize at the exact same critical value of $\lambda$, a staggered potential self-interferes differently across the energy spectrum causing different energy states to localize at different values of $\lambda$. Though with increase in disorder strength, the QP first weakens the coherent dimerization to produce localization, the spectrum does not localize collectively for intermediate $\lambda$ values as the competition with the staggered hopping redistributes the wave-function amplitudes over larger regions of the lattice, giving rise to a partial recovery of extended character reflected by the simultaneous finite values of the averaged $<NPR>$ and $<IPR>$. It also results in splitting of the SPME into two different energies in the energy spectrum, freeing certain energy bands to behave as extended states as the QP strengthens. But then for sufficiently large QP strength, the competition weakens and localization is restored again.
\\


{\section{Different Hopping Periodicities}}
\begin{figure}
   \begin{picture}(100,100)
     \put(-200,20){
       \includegraphics[width=.32\linewidth,height= 1.3 in]{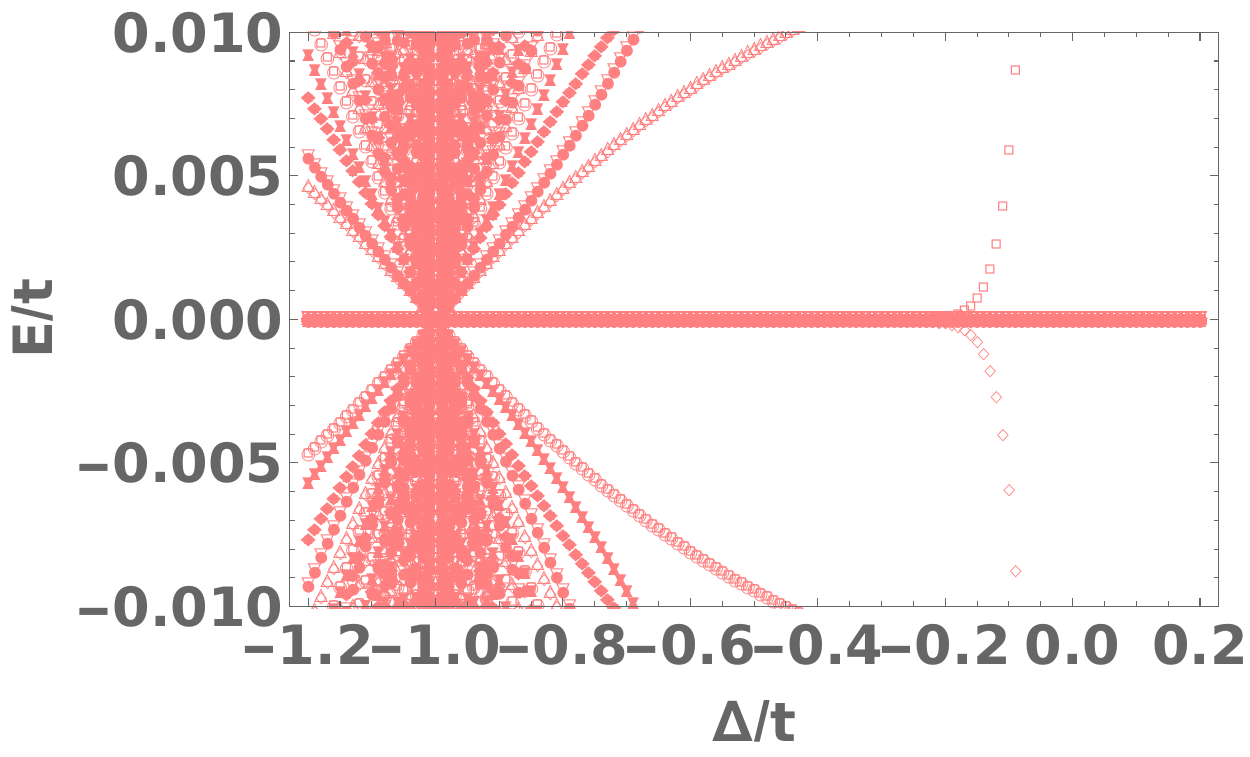}
     \includegraphics[width=.32\linewidth,height= 1.3 in]{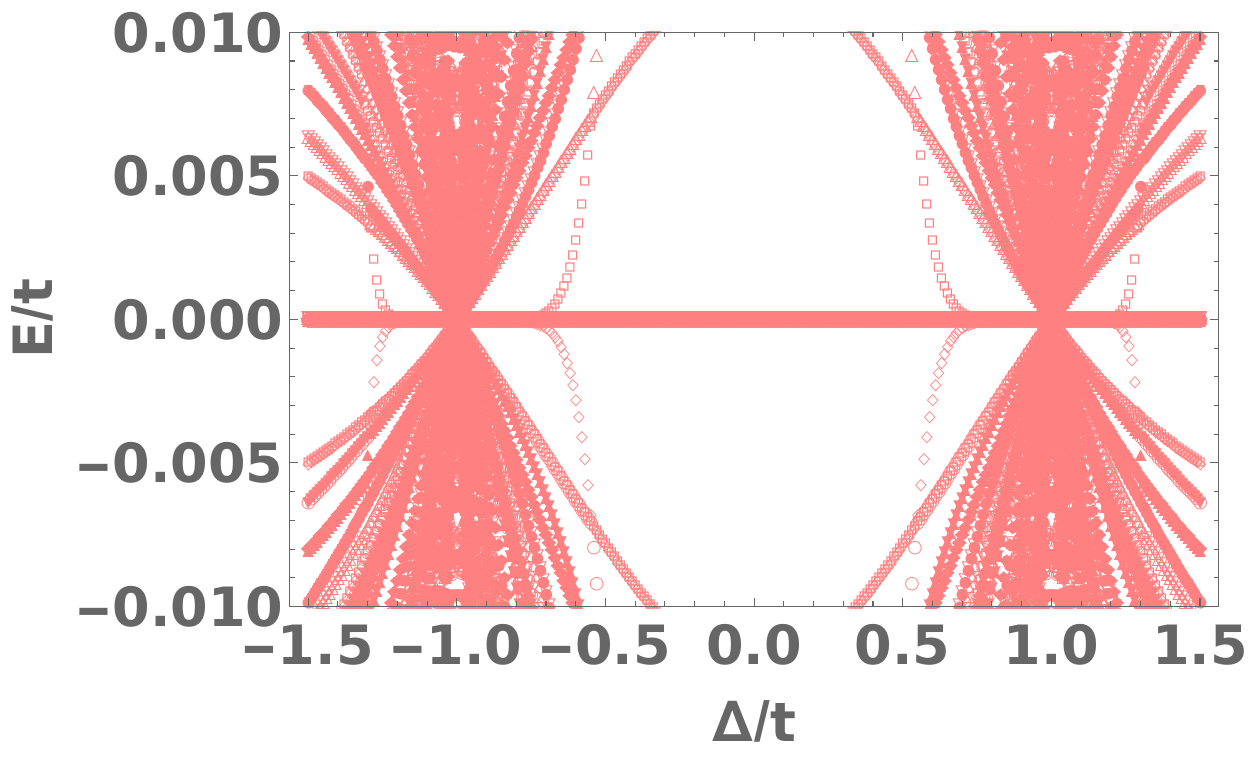}
  \includegraphics[width=.32\linewidth,height=1.3 in]{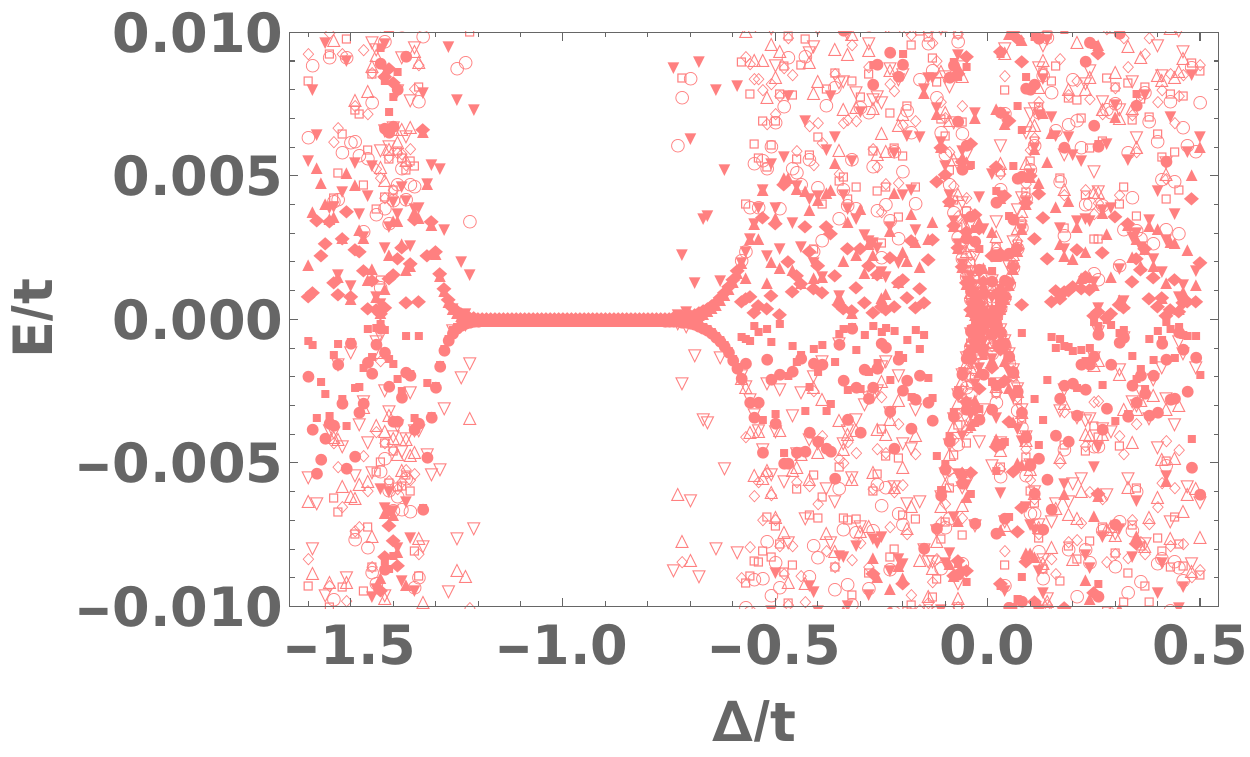}}\hskip .4 in
          \put(-95,90){(a)}
     \put(60,90){(b)}
     \put(205,90){(c)}
   \end{picture}\\\vskip .1 in
   \begin{picture}(100,100)
     \put(-180,30){
     \includegraphics[width=.3\linewidth,height= 1.3 in]{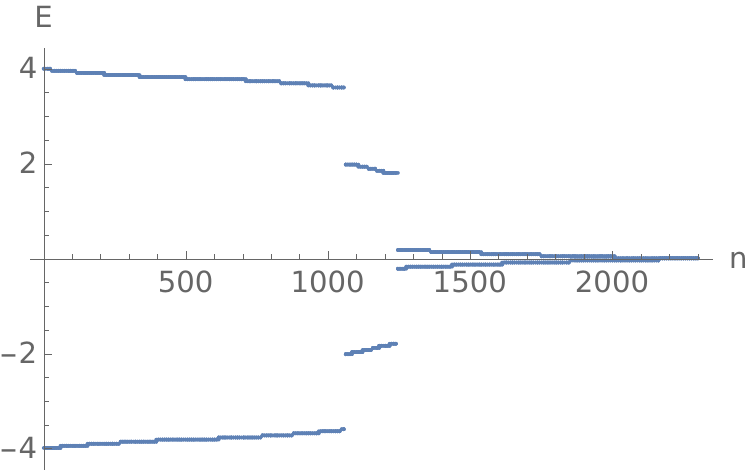}
     \includegraphics[width=.3\linewidth,height= 1.3 in]{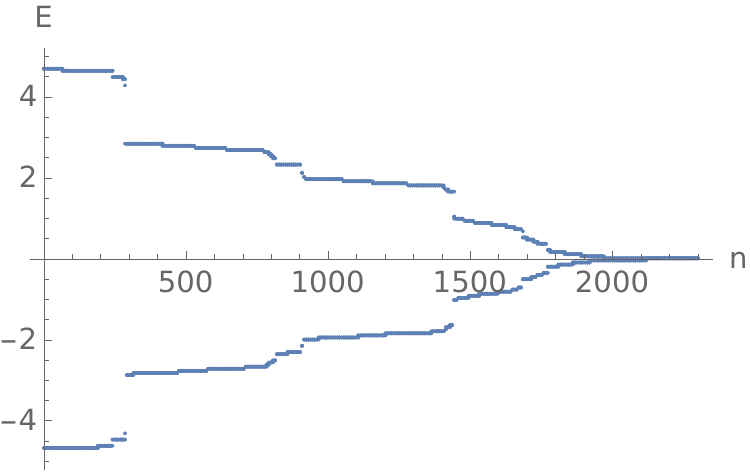}
  \includegraphics[width=.3\linewidth,height=1.3 in]{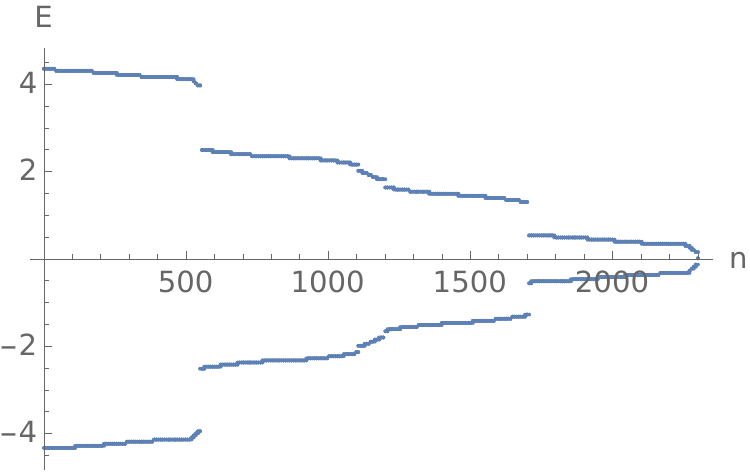}}\hskip .4 in
     \put(-95,100){(d)}
     \put(60,100){(e)}
     \put(205,100){(f)}
   \end{picture}
   \vskip -.4 in
   \caption{{ $2D$ SSH model low energy spectra for a range of $\Delta/t$ (top panel) and energy eigenvalue E vs. eigenstate index n for $\Delta/t= -0.9$ (bottom panel) in a $48\times 48$ lattice.} Here, (a,d) $\theta_{x}=\theta_y=\pi$, (b,e) $\theta_x=\theta_y=\pi/2$ and (c,f) $\theta_x=\pi,~\theta_y=\pi/2$.} 
\label{dispersion3}
\end{figure}
\begin{figure}
     \includegraphics[width=.9\linewidth,height= 1.5 in]{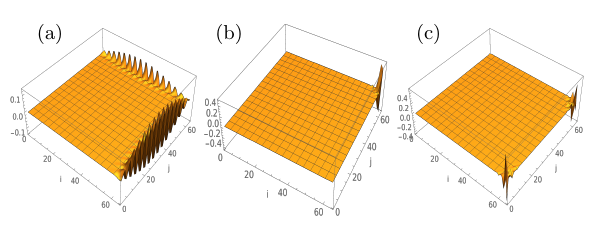}
   \caption{Typical (a) type-III, (b) type-IV and (c) type-V in-gap states of a $2D$ SSH model in a $64\times 64$ lattice for $\theta_x=\theta_y=\pi/2$. } 
\label{dispersion4}
\end{figure}

Next, we briefly study the case of $(\theta_{x},\theta_{y})=(\pi/2,\pi/2)$ and $(\pi,\pi/2)$ to probe its spectra and also the outcome of introducing domain walls in it. In the first case, the spectra remains independent of the sign of $\Delta/t$, unlike that for $\theta=\pi$, while in the latter the spectra is very asymmetric with respect to $\Delta/t$ (see Ref.\cite{kar}). {The topological phases are again associated with vector Zak phases of $Z(\pi,\pi)$, as simple extensions of what we found in 1D hopping modulated SSH chain\cite{mandal}. Notice that the two components of the vector Zak phase in these cases remain the same as long as we keep same hopping parameter $t$ and modulation amplitude $\Delta$ along the $x$ and $y$ directions.} In Fig.\ref{dispersion3}, the numerical energy spectra for small energies as well as spectra at $\Delta/t=-0.9$ in a $48\times48$ lattice are shown for the two cases as well as for the $(\theta_x,\theta_y)=(\pi,\pi)$ case {for the sake of comparison.}

{\subsection{Case I: $\theta_x=\theta_y=\frac{\pi}{2}$}}
For $(\theta_{x},\theta_{y})=(\pi/2,\pi/2)$, {a $48\times48$ system shows zero energy BICs specifically for $1.4\gtrsim|\Delta/t|\gtrsim0.8$ manifesting its topological regime. Fig.\ref{dispersion3}(b) demonstrates the variation of low energy spectra with $\Delta/t$ where  two nonzero energy bulk modes can be seen turning into ZES at around $|\Delta/t|\sim0.8,1.4$.}
Similar to the $(\pi,\pi)$ case, here also the spectra from a $L\times L$ lattice contains $L+2~(L)$ number of ZES within the topological (trivial) regime (but away from the dimerization-free point $\Delta=0$).

Earlier we have mentioned that the in-gap states in case of $\theta=\pi$, as seen in Fig.\ref{dispersion3}(d), are topological characterized by nonzero 2D Zak phases\cite{liu}. {For $(\theta_x,\theta_y)=(\frac{\pi}{2},\frac{\pi}{2})$ as well, we obtain topological in-gap states - only more in number this time due to further increased periodicities in the hopping. Hence more gaps open in the spectrum (see Fig.\ref{dispersion3}(e)),} as also witnessed in the equivalent 1D problem\cite{kar}.
\begin{table*}
\parbox{.82\linewidth}{
\centering
\begin{tabular}{  p{3cm}| p{3cm}| p{3cm}  |p{3cm} |p{3cm} } 
 \hline
 Periodicity in both directions & Vertical edge states & Horizontal edge states & Both Horizontal and vertical edge states  & Corner states \\ [0.65ex] 
 \hline
 $\theta_{x}=\pi,\theta_{y}=\pi$ & Yes & Yes  & Yes &Yes \\ [1.2ex] 
 \hline
  $\theta_{x}=\pi/2,\theta_{y}=\pi/2$ & Yes  & Yes & Yes &Yes \\ [1.2ex] 
 \hline
  $\theta_{x}=\pi,\theta_{y}=\pi/2$ & Yes & No & No & Yes \\ [1.2ex] 
 \hline
  $\theta_{x}=\pi/2,\theta_{y}=\pi$ & No & Yes & No & Yes\\[1.2ex] 
 \hline
\end{tabular}
\caption{Existence of corner, horizontal edge, and vertical edge states for a few possible combinations of hopping periodicities in $x$ and $y$ directions.}
\label{table1}}
\end{table*}
Other than type I and type II in-gap states observed in the $\theta=\pi$ case, here one can additionally witness new kinds of in-gap states {(See Table 1 for details)} where edge modes are localized in a single $x$ and $y$-edge (type-III) or in single or multiple corners (type IV and V respectively). Examples of such states are shown in Fig.\ref{dispersion4}.\\

{\subsection{Case II: $\theta_x=\pi,~\theta_y=\frac{\pi}{2}$}}

The case for $\theta_{x}=\pi$ and $\theta_{y}=\pi/2$ provides noteworthy modifications in the spectra and topology.
 {For anisotropic hoping modulations such as this one (see Fig.\ref{dispersion3}(c)), only 4 ZES can be obtained in the topological phase - which are all boundary modes.
Particularly, these are obtained only for $-1.2\lesssim\Delta/t\lesssim -0.8$ in the spectra from a $48\times48$ size lattice. There no more remains degenerate bulk modes at zero energy (see Fig.\ref{dispersion3}(f)). Thus}
unlike in the previous cases of $\theta_{x}=\theta_{y}=\pi~\&~\pi/2$, here topological ZES are no longer buried within the continuous bulk spectra\cite{Benalcazar,wei}, but are separated from them via a band gap.
Our results show that in a $2D$ SSH model, engagement of next-nearest-neighbor couplings is not the only way to create a band gap within the continuous band structure\cite{wei}. One can also open such a band gap by considering different periodicity for the hopping modulation in the two directions as we considered in this case.
Interestingly, the zero energy modes at $\Delta/t=-0.8$ are confined to all the corners in contrary to the ZES for $\Delta/t=-0.9$ where corner modes are localized at the two adjacent corners (specifically, the corners containing $y$-edge) of the $2D$ lattice.
Usually the 2D SSH model Hamiltonian with $\theta_x=\theta_y=\pi$ can be described as the sum of the Hamiltonians
of parallel SSH chains in the x and y directions because its matrix representation can be shown to be Kronecker product of tridiagonal SSH chain matrix with the identity matrix (symbolically $H_{SSH}^{2D}=H_{SSH}^{1D}\otimes\hat{I_2}+\hat{I_2}\otimes H_{SSH}^{1D}$ and $E_{SSH}^{2D}=E_{SSH}^{1D}(k_x)+E_{SSH}^{1D}(k_y)$) generating 4 bulk bands in the spectrum\cite{dias2}. {Thus it gives zero modes whenever $E_{SSH}^{1D}(k_x)=-E_{SSH}^{1D}(k_y)$ which appears due to chiral symmetry in the system and is manifested at $|k_x|=|k_y|$ producing continuum of zero energy bulk states. But for different dimerization along $\hat x$ and $\hat y$, $e.g.,$ for the case of $\theta_{x}=\pi$ and $\theta_{y}=\pi/2$, SSH chains along two directions feature dispersions corresponding to different hopping parameters and the condition  $E_{SSH}^{1D}(k_x)=-E_{SSH}^{1D}(k_y)$ is never met barring the spectrum of any bulk ZES.}

{Like in isotropic cases, here also one can consider the vector Zak phase to be the topological invariant for the system. But unlike the isotropic case, here one can witness both same and different combination of $Z_x$ and $Z_y$ components in the topological regime in a $\Delta-t$ parameter space.}

{Its also worth mentioning in this case of $\theta_x=\pi,~\theta_y=\frac{\pi}{2}$, that the topological in-gap modes come out to be vertical edge modes.}\\


\begin{figure}
      \includegraphics[width=.9\linewidth,height= 2.5 in]{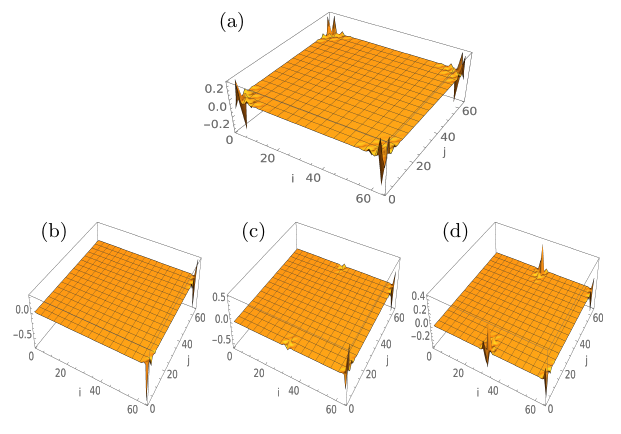}
\caption{(a) Typical zero energy BICs of a finite $2D$ SSH model in a $64\times 64$ lattice for $\theta_{x}=\pi/2,~\theta_{y}=\pi/2$, $i_0=j_0=32$ and for $\Delta/t= -0.9$. (b-d) Typical ZES in case of $\theta_{x}=\pi,~\theta_{y}=\pi/2$ in absence (b) and presence (c,d) of DWs.} 
\label{dwpb2pb2}
\end{figure}

{\subsection{Case III: $\theta_y=\pi,~\theta_x=\frac{\pi}{2}$}}

Topological ZES and in-gap modes for $\theta_{x}=\pi/2, \theta_{y}=\pi$ show opposite trend as compared to that for $\theta_{x}=\pi, \theta_{y}=\pi/2$.
Unlike for $\theta_{x}=\pi, \theta_{y}=\pi/2$, this case features only horizontal edge modes. The appearance of topological in-gap states and corner states for various combinations of hopping periodicity in both directions is summarized in Table.\ref{table1}.\\

{\subsection{DWs in presence of periodic modulations}}

Finally we examine the engagement of the DW on these periodically modulated models. Particularly, we introduce static DWs via Eq.\ref{dw1} in these systems. We find that the spectral symmetry with respect to $\Delta$ (or $d_0$, in this scenario) in the $(\pi/2,\pi/2)$ case is lost due to the presence of antiphase domain walls. Also like in 1D\cite{mandal}, we witness corner modes with no localization at the DW center as the new zero energy BICs (see Fig.\ref{dwpb2pb2}(a)).
Unlike the previous case of $\theta=\pi$, here the BICs exhibit finite amplitudes at all the corners. The four  ZES (and no BIC anymore) for $\theta_{x}=\pi$ and $\theta_{y}=\pi/2$ are found localized (for $\Delta/t=-0.9$) at two corners at single $x$ edge (see Fig.\ref{dwpb2pb2}{(b)}). However on bringing DWs within the lattice, those ZES show additional peaks at the $x$ positions same as that of the DW center (see Fig.\ref{dwpb2pb2}(c,d)). \\

\section{Summary and conclusions}

In this paper, we have considered a 2D SSH model with periodic hopping modulations in presence of a series of domain walls and a quasi-periodic disorder. Firstly, the SSH model spectra is analyzed for both periodic and open boundaries in a square lattice. The nature of different types of the zero energy states and topological in-gap states are discussed. Then we introduce a series of domain walls in such a system first along two orthogonal lines usually intersecting at the center of the lattice and then along a circle imitating a vortex within the lattice. In the first case, we find the {nonzero energy} in-gap states to populate along the edges and the DW lines and the ZES to show corner peaks or peaks at the intersection of the DW lines. In the second case, however, the in-gap states remain localized only at the center of the circle of DWs while the ZES can show states which localizes around that center in a symmetric fashion in the square lattice.

A diagonal quasiperiodic disorder can provide localization to the eigenstates consistently as we show using the $\langle IPR\rangle$ and $\langle NPR\rangle$ plots.
Moreover like in 1D, we also find that smart choices of quasiperiodic potentials can produce nontrivial behavior where a re-entrance in localizations, as recognized from the $\langle NPR\rangle$ results, can be observed. Though the observed reentrant $<NPR>$ peak is not strong enough, our findings indicate the possibility of realizing such exotic behavior in 2D quasiperiodic models which has not yet been reported elsewhere and thus leave the task open to the community to look for more prominent signatures of the same.

Then considering different hopping periodicities like $(\theta_{x},\theta_{y})=(\pi/2,\pi/2)$, $(\pi,\pi/2)$ and $(\pi/2,\pi)$, we first show how this modulation introduces additional gaps in the energy spectrum and how the number of ZES gets affected. We also display new kinds of in-gap states observed in case of $(\theta_{x},\theta_{y})=(\pi/2,\pi/2)$. Finally, we introduce domain walls in these periodically hopping modulated systems and find out how the zero energy modes {resemble} in presence of such defects.

Our findings can add important leads in the condensed matter research and innovation for its possible application in the field of topological quantum computations and quantum transport. {As per future prospects, our non-interacting model features can be fed into supervised learning algorithms for realizing topological phases of different interacting topological insulators\cite{scipost}. In a 1D 
SSH chain, topological transitions can be simulated digitally and detected using trapped ion quantum computers\cite{quan-comp}. We can follow the same in our 2D system as well with a guidance of phase details from the present work. Our reported re-entrance localization phenomena which is rather rare to occur in a 2D system, also deserves further investigations for the novelty it points to.
}

One can always verify the results {shown in this work} in a cold atom set-up within optical lattices\cite{xie}. The techniques of controlling nearest-neighbor couplings have already been developed in synthetic quantum materials like photonic and acoustic crystals\cite{li}.
Experimental realization of 2D SSH model in acoustic systems resulted in manifestation of topological edge waves and HOTI as found in a simple network of air channels\cite{zheng}.  In a 2D SSH model on a square lattice we find higher order topological corner states at zero energy. As BICs, they withstand localized natures even being degenerate with many bulk modes and thereby represent highly coherent states immuned to leaking into the bulk\cite{weimann}. This can be very useful in providing stable quantum information storage in 2D architectures. In particular, such confinement with quadrupole topological charge\cite{hughes} in a 2D SSH like model with complex hopping amplitudes\cite{pra104} can produce topological qubits encoding quantum information in superconducting circuit systems\cite{pra104}. Furthermore, by adiabatically tuning the hopping parameters, one can move a corner state along the boundary to other corners and thus braid the localized modes to perform quantum computing operations\cite{braiding}.
Engineering specific DW configurations, like that adopted in our work, one can create topological channels anywhere within the bulk of the material that can create reconfigurable topological wires or corner qubits - not just restricted to the physical perimeters\cite{noh,he}. The non-monotonic phase behavior in a reentrance localization as reported in our work can offer a novel "switching" mechanism for quantum transport\cite{s-basu}. Also an hopping anisotropy allows for the selective enhancement of boundary modes by isolating the topological ZES from the gapped states. It reduces the dimensionality and hence the mode volumes of the boundary modes. This helps in designing high sensitivity sensors or lasers with tiny mode volumes\cite{parto}.

\section*{Acknowledgements}
The authors thank F. Liu, {W. Benalcazar}, S. Basu, S. Maiti and S. Roy for important feedbacks. Thanks also to Rajdip Banerjee for his help in plotting the cartoons of $\delta$ profiles in Fig.3,5. This work is financially supported by DST-SERB (now called ANRF), Government of India via grant no. CRG/2022/002781.

\appendix

  \section{Energy eigenstates, Parity and Band Inversion}
  \label{ap1}
  The Bloch Hamiltonian for the 2D SSH model given by Eq.\ref{1} is a $4\times4$ Hermitian matrix $H_k$ with nonzero elements $H_k(1,2)=H_k(3,4)=t-\Delta+(t+\Delta)e^{-ik_x}$, $H_k(1,3)=H_k(2,4)=t-\Delta+(t+\Delta)e^{-ik_y}$ and their conjugates\cite{liu,obana,kar}. {One can see that it can be split into two parts, each being an outer product of a 1D SSH Hamiltonian matrix and an identity matrix, as mentioned in Sec.IIB, with $\{H_x(k_x),H_y(k_y)\}=$
    \begin{equation}
  \left\{\left[\begin{array}{cc}0&H_k(1,2)\\H_k^\star(1,2)&0\end{array}\right],\left[\begin{array}{cc}0&H_k(1,3)\\H_k^\star(1,3)&0\end{array}\right]\right\}.
\end{equation} } The spectrum {of such 2D SSH Hamiltonian} is given by Eq.\ref{4} and the eigenstates come out to be
\begin{align}
  |u_j(k)\rangle=\frac{1}{2}\left(\begin{array}{c}
      1\\ s_1(j)e^{-i\phi(k_x)}\\ s_2(j)c^{-i\phi(k_y)}\\ s_1(j)s_2(j)e^{-i[\phi(k_x)+\phi(k_y)]}\end{array}\right)
\end{align}
where $[s_1(j),s_2(j)]~$= [-1,-1], [-1,+1], [+1,-1] and [+1,+1] for $j=$ 1, 2, 3 and 4 respectively and $\phi(k_l)=tan^{-1}\frac{(t-\Delta)\sin k_l}{t+\Delta+(t-\Delta)\cos k_l}$.
As mentioned earlier, this shows a chiral symmetric spectrum with double degeneracy at zero energy for $|k_x|=|k_y|$. Notice that due to $C_{4v}$ symmetry, $H_k$ remains invariant under a $\pm\pi/2$ rotation about the $\hat z$ axis. A $C_2$ rotational symmetry makes $E(k_x,k_y)$ a function of only the magnitude of Bloch vectors $k_k$ and $k_y$. A $C_4$ rotational symmetry imposes additional restriction such that one can write $E(k_x,k_y)=f(|k_x|,|k_y|)-f(|k_y|,|k_x|)+C,~f$ and $C$ being a function of $k_x$ and $k_y$ and a constant respectively. Such conditions, in addition to the chiral symmetry brings in zero dispersions along the nodal directions $|k_x|=|k_y|$. One can easily check that $H_k$, being chiral symmetric, can be unitary transformed (via swapping of a row followed by a column) to an anti-diagonal form\cite{Benalcazar}: 
\begin{equation}
  H_k\rightarrow\left[\begin{array}{cc}0&Q\\Q^\dagger&0\end{array}\right] \rm{where~}
  Q=\left[\begin{array}{cc}
  H_k(1,2)&H_k(1,3)\\
  H_k(1,3)^\star&H_k(1,2)^\star
  \end{array}\right].
\end{equation}
But for $|k_x|=|k_y|$, we get $Det[Q]=0$ implying a pair of zero energy states (if not four zero energy states that occur when $Q$ becomes a null matrix at ${\bf k}=(\pm\pi,\pm\pi)$ at TQPT) in the spectrum.

In a 2D Su-Schrieffer-Heeger (SSH) model, the parity of energy eigenstates is typically determined by examining the inversion symmetry of the Bloch Hamiltonian at specific high-symmetry points in the Brillouin zone. The inversion or parity operator $\mathcal P$ acts by inverting the crystal momenta (${\bf k\rightarrow -k}$) and transforming the internal degrees of freedom. For a 4-band model like ours, this is typically given as
\begin{equation}
   \mathcal P=\tau_x\otimes\sigma_x=
\begin{pmatrix}
0 & 0 & 0 & 1 \\
0 & 0 & 1 & 0 \\
0 & 1 & 0 & 0 \\
1 & 0 & 0 & 0
\end{pmatrix}.
\end{equation}
The SSH Hamiltonian obeys parity symmetry at the high symmetry points $\Gamma=(0,0)$, $M=(\pi,\pi)$ and $X(Y)=(\pi,0)((0,\pi))$.
Notice that for $\Delta/t>0$, $\mathcal P|u_j(k)\rangle=(-1)^{s_1(j)+s_2(j)}|u_j(k)\rangle$ at these points. However for $\Delta/t<0$ ($i.e.,$ the topological regime), $\mathcal P|u_j(k)>$ changes sign at the $X(Y)$ point. In fact, though $\phi(k_l=0)=0$, $lim_{k_l\rightarrow\pi^-}\phi(k_l)=tan^{-1}\frac{(t-\Delta)\sin\pi^-}{-2\Delta\cos\pi^-}=0(\pi)$ for $\Delta/t>0~(~\Delta/t< 0~)$. This makes $\mathcal P|u_j(k)\rangle=-(-1)^{s_1(j)+s_2(j)}|u_j(k)\rangle$ for $\Delta/t<0$ at the X and Y point. Thus it indicates parity or band inversions. This is a signature of the TQPT at $\Delta=0$. There gaplessness occur also at the $M$ points but without any parity inversions. So this represents a trivial band crossing at $\Delta=0$.

{Zak phase in a SSH chain possess a nonzero value of $\pi$ in the topological regime. Similarly,} vector Zak phases, that appear as topological invariant in this problem, take quantized nonzero values {in its two components} in the topological regime for $\Delta/t<0$. Following Eq. \ref{zak}, one can verify the same. For example, the $x$-component of the vector Zak phase, $Z_x$, comes out to be
\begin{align}\label{zx}
  Z_x&=\sum_{j=1}^{N_{occ}}\int_{-\pi}^\pi\langle u_j(k)|i\frac{\partial}{\partial k_x}|u_j(k)\rangle dk_x=\sum_{j=1}^{N_{occ}}\frac{1}{4}\int_0^\pi [1-\frac{2t\Delta}{(t^2-\Delta^2)\cos k_x+(t^2+\Delta^2)}] dk_x=\frac{\pi}{2}[1-sgn(t*\Delta)]
  \nonumber\\&=\begin{cases} 0 & (\Delta/t>0)\\\pi & (\Delta/t<0)\end{cases}
\end{align}
for $N_{occ}=1$ (a quarter-filled system). Notice that the $k_y$ dependence of $|u_j(k)\rangle$ does not show up in $Z_x$ due to the scalar product within the integral. Similar calculations can be performed for $Z_y$ yielding same result.
This quantized Zak-phase also represents charge displacements/dipole moments\cite{Benalcazar} caused by modulated hopping in the Hamiltonian and are related to the polarization $P$ via  relation $P_l=Z_l/2\pi$. So in the topological regime, $P_l$ becomes fractional.

\section{Wannier function and polarization}
\label{ap2}
A Wannier function is a localized wave function used to describe electrons in a crystalline solid. While the standard Bloch waves ($\psi_{nk}(r)=e^{ik.r}u_{nk}$) are spread out across the entire crystal, Wannier functions are constructed as the Fourier transform of Bloch functions:$$w_n(\mathbf{r} - \mathbf{R}) = \frac{V}{(2\pi)^3} \int_{BZ}  \, e^{i\mathbf{k} \cdot (\mathbf{r-R})} u_{n\mathbf{k}}~d\mathbf{k}$$ where $n$, $R$ and $V$ denote the band index, lattice vector and volume of the crystal respectively.
The Wannier Center is the average position of an electron described by a Wannier function and thus defined as:$$\bar{\mathbf{r}}_n = \int d\mathbf{r} \, \mathbf{r} |w_n(\mathbf{r})|^2$$.
In the modern theory of polarization\cite{hughes}, The displacement of Wannier centers relative to the ionic cores is directly related to the electric polarization of a crystalline solid.
In topological materials, the way Wannier centers shift as a function of momentum can be used to identify the topological invariants.


\begin{thebibliography}{00}
\bibitem{kane}M. Z. Hasan and C. L. Kane, Colloquium: Topological insulators, {Rev. Mod. Phys. {\bf 82}, 3045 (2010).}
\bibitem{zhang1}X.-L. Qi and S.-C. Zhang, Topological insulators and superconductors, {Rev. Mod. Phys. {\bf 83}, 1057 (2011).}
\bibitem{hughes}W. A. Benalcazar, B. A. Bernevig, and T. L. Hughes, {Science {\bf 357}, 61 (2017).}
\bibitem{edge}B.-Y. Xie $et.~al.$, {Phys. Rev. B {\bf 98}, 205147 (2018).}
{\bibitem{second} M Geier $et.~al.$, {Phys. Rev. B {\bf 97}, 205135 (2018).}}
\bibitem{su}W.P. Su, J.R. Schrieffer, and A.J. Heeger, Solitons in Polyacetylene, {Phys. Rev. Lett. {\bf 42}, 1698 (1979).}
\bibitem{heeger}W.-P. Su, J. R. Schrieffer, and A. J. Heeger, Soliton excitations in polyacetylene, {Physical Review B {\bf 22}, 2099 (1980).}
\bibitem{rebbi} R. Jackiw and C. Rebbi, Solitons with fermion number $\frac{1}{2}$, {Phys. Rev. D {\bf 13}, 3398 (1976).}
\bibitem{lee}E. Lee, A. Furusaki, and B.-J. Yang, Fractional charge bound to a vortex in two-dimensional topological crystalline insulators, {Phys. Rev. B {\bf 101}, 241109(R) (2020).}
\bibitem{liu}F. Liu, and K. Wakabayashi, Novel topological phase with a zero Berry curvature, \textcolor{blue}{Phys Rev Lett. {\bf 118}, 076803 (2017).}
  
\bibitem{kar}S. Kar, Edge state behavior in a Su–Schrieffer–Heeger like model with periodically modulated hopping, {J. Phys.: Condens. Matter, {\bf 36}, 065301 (2024).}
\bibitem{karc}S. Kar, Corrigendum: Edge state behavior in a Su–Schrieffer–Heeger like model with periodically modulated hopping (2024 J. Phys.: Condens. Matter 36, 065301), {J. Phys.: Condens. Matter, {\bf 37}, 279501 (2025).}
\bibitem{hou}C.-Yu Hou, C. Chamon, and C. Mudry, Electron Fractionalization in Two-Dimensional Graphenelike Structures,  {PRL {\bf 98}, 186809 (2007).}
\bibitem{aubry} S. Aubry, G. André, Analyticity breaking and Anderson localization in incommensurate lattices, {Ann. Israel Phys. Soc. {\bf 3}, 18 (1980).}
\bibitem{ander} P. W. Anderson, Absence of Diffusion in Certain Random Lattices
, {Phys. Rev.{\bf 109}, 1492 (1958).}
\bibitem{s-basu} S. Roy, T. Mishra, B. Tanatar, and S. Basu, Reentrant Localization Transition in a Quasiperiodic Chain, {Phys. Rev. Lett. {\bf 126}, 106803 (2021).}
\bibitem{chiral}A. Maiellaro, and R. Citro, Topological Edge States of a Majorana BBH Model, {Condens. Matter {\bf 6}, 2 (2021).} 
\bibitem{Benalcazar}W. A. Benalcazar and A. Cerjan, Bound states in the continuum of higher-order topological insulators, {Phys. Rev. B {\bf 101}, 161116(R) (2020).}
\bibitem{obana}D. Obana, F. Liu and K. Wakabayashi, Topological edge states in the Su-Schrieffer-Heeger model, {Phys. Rev. B {\bf 100}, 075437 (2019).}
\bibitem{mandal}S. Mandal and S. Kar, Topological solitons in a Su-Schrieffer-Heeger chain with periodic hopping modulation, domain wall, and disorder, {Phys. Rev. B {\bf 109},
195124 (2024).}
\bibitem{nhmandal}S. Mandal, and S. Kar, Topology and PTSymmetry in a Non-Hermitian Su-Schrieffer-Heeger Chain with Periodic Hopping Modulation, {J. Phys.: Condens. Matter {\bf 37} 095602 (2025).}
\bibitem{wei}M.-S. Wei, M.-J. Liao, C. Wang, C. Zhu, Y. Yang, and J. Xu, Topological laser with higher-order corner states in the 2-dimensional Su-Schrieffer-Heeger model, {Opt. Express {\bf 31}, 3427-3440 (2023).}
\bibitem{ma}H. Ma, Z. Zhang, P.-H. Fu, J. Wu, and X.-L. Yu, Electronic and topological properties of extended two-dimensional Su-Schrieffer-Heeger models and realization of flat edge bands, {Phys. Rev. B {\bf 106}, 245109 (2022).}
\bibitem{xu}X.-W. Xu, Y.-Z. Li, Z.-F. Liu, and A.-X. Chen, General bounded corner states in the two-dimensional Su-Schrieffer-Heeger model with intracellular next-nearest-neighbor hopping, {Phys. Rev. A {\bf 101}, 063839 (2020).}
\bibitem{dong}W.-Jin Zhang, H.-Chang Mo, W.-Jie Chen, X.-Dong Chen, and J.-Wen Dong, Observation of gapless corner modes of photonic crystal slabs in synthetic translation dimensions, {Photon. Res. {\bf 12}, 444-455 (2024).}
\bibitem{dias}G. Pelegr\'{i}, A. M. Marques, V. Ahufinger, J. Mompart, and R. G. Dias, Second-order topological corner states with ultracold atoms carrying orbital angular momentum in optical lattices, {Phys. Rev. B {\bf 100}, 205109 (2019).}
\bibitem{dias2}R. G. Dias, L. Madail, A. Lykholat, R. Andrade, and A. M. Marques, Topological wave equation eigenmodes in continuous 2D periodic geometries, {Eur. J. Phys. {\bf 45} 045801 (2024).}
\bibitem{benal2}  W. A. Benalcazar, T. Li, and T. L. Hughes, Quantization of fractional corner charge in Cn symmetric highe-order topological crystalline insulators, {Phys. Rev. B {\bf 99} 245151 (2019).}
  \bibitem{jackiw}R. Jackiw, Fractional and Majorana fermions: the physics of zero-energy modes, {Phys. Scr. {\bf T146} 014005 (2012).}
\bibitem{chen}X. Zhou, Z. Wang, H. Chen, Topological Aspects of Dirac Fermions in a Kagom\'{e} Lattice, { arXiv:2412.04010v2 [cond-mat.str-el}.

  \bibitem{domain}M. Scollon and M. P. Kennett, ``Persistence of chirality in the Su-Schrieffer-Heeger model in the presence of on-site disorder", 
    {Phys. Rev. B {\bf 101}, 144204 (2020).}
\bibitem{quasicrystal} Jean-Noël Fuchs, and Julien Vidal, Hofstadter butterfly of a quasicrystal, {Phys. Rev. B {\bf 94}, 205437 (2016).}    
\bibitem{prb101-014205} A. Szabo, and U. Schneider, Mixed spectra and partially extended states in a two-dimensional quasiperiodic model, {Phys. Rev. B {\bf 101}, 014205 (2020).}
{\bibitem{scipost} P. Molignini $et.~al.$, ``A supervised learning algorithm for interacting topological insulators based on local curvature'', Scipost Phys. {\bf 11}, 073 (2021).}
 \bibitem{quan-comp} Q. Xie $et.~al.$, ``Digital quantum simulation of the Su-Schrieffer-Heeger model using a parameterized quantum circuit", {arXiv:2504.07499 (2025).}

\bibitem{xie}D. Xie $et.~al.$, ``Topological characterizations of an extended Su–Schrieffer–Heeger model", {njp Quant. Inf. {\bf 5}, 55 (2019).}
  \bibitem{li}C.-A. Li, ``Topological states in two-dimensional Su–Schrieffer–Heeger models", {Front. Phys. {\bf 10}, 861242 (2022).}
\bibitem{zheng}L.-Y. Zheng, V. Achilleos, O. Richoux, G. Theocharis, and V. Pagneux, Observation of Edge Waves in a Two-Dimensional Su-Schrieffer-Heeger Acoustic Network, {Phys. Rev. Applied {\bf 12}, 034014(2019).}
    \bibitem{weimann} S. Weimann $et.~al.$, Topologically protected bound states in photonic parity-time-symmetric crystals, {Nature Materials {\bf 16(4)}, 433 (2017).}

 \bibitem{pra104} C. Wu $et.~al.$, Dynamical characterization of quadrupole topological phases in superconducting circuits, {Phys. Rev. A {\bf 104}, 022601 (2021).}
    { \bibitem{braiding} T. E. Pahomi $et.~al.$, Braiding Majorana corner modes in a second-order topological superconductor, {Phys. Rev. Res. {\bf 2(4)}, 032068(R) (2020).} }
    
 \bibitem{noh} J. Noh $et.~al.$, Topological protection of light in the continuum, {Nature Photonics {\bf 104}, 022601 (2018).}
 \bibitem{he} J. He $et.~al.$, Mode engineering in reconfigurable fractal topological circuits, {Phys. Rev. B {\bf 109}, 235406 (2024).}
   \bibitem{parto} M. Parto $et.~al.$, Edge-Mode Lasing in 1D Topological Lattices, {Phys. Rev. Lett. {\bf 120}, 113901 (2018).}


\end{thebibliography}
\end{document}